\documentclass[sigconf, screen]{acmart}

\AtBeginDocument{%
  \providecommand\BibTeX{{%
    \normalfont B\kern-0.5em{\scshape i\kern-0.25em b}\kern-0.8em\TeX}}}

\copyrightyear{2025}
\acmYear{2025}
\setcopyright{cc}
\setcctype{by}
\acmConference[CHI '25]{CHI Conference on Human Factors in Computing Systems}{April 26-May 1, 2025}{Yokohama, Japan}
\acmBooktitle{CHI Conference on Human Factors in Computing Systems (CHI '25), April 26-May 1, 2025, Yokohama, Japan}\acmDOI{10.1145/3706598.3714187}
\acmISBN{979-8-4007-1394-1/25/04}

\usepackage{subcaption} 

\usepackage{url}
\makeatletter
\g@addto@macro{\UrlBreaks}{\UrlOrds}
\makeatother

\usepackage{multirow} 
\usepackage{color} 
\usepackage{enumitem} 

\usepackage{xspace}
\usepackage{xcolor, colortbl}

\usepackage{arydshln}

\newcommand{\m}{\textit{M=}}

\newcommand{\sd}{\textit{SD=}}

\usepackage{tcolorbox}
\usepackage{hyperref}

\MakeRobust{\ref}

\makeatletter
\newcommand{\labeltext}[2]{%
  \@bsphack
  \csname phantomsection\endcsname 
  \def\@currentlabel{#1}{\label{#2}}%
  \@esphack
}
\makeatother

\newtcbox{\highlight}[1][magenta]{on line, arc=0pt,colback=#1!10!white,colframe=#1!50!black, before upper={\rule[-3pt]{0pt}{10pt}},boxrule=1pt, boxsep=0pt,left=4pt,right=3pt,top=2pt,bottom=1pt}

\newcommand{\researchq}[2]{
    \noindent 
    \highlight{\textbf{\texttt{\textcolor{gray}{#1}}}} 
    \textit{#2} 
    \labeltext{\highlight{\textbf{\texttt{\textcolor{gray}{#1}}}} \textit{#2}} 
    {rq:#1} 
}

\newcommand{\notextrefrq}[1]{\hyperref[rq:#1]{\highlight{\textbf{\texttt{\textcolor{gray}{#1}}}}}}

\def\plainkeywords{External communication; Autonomous vehicles; Pedestrian Behavior; eHMI.}

\begin{document}

\title{Improving External Communication of Automated Vehicles Using Bayesian Optimization}

\author{Mark Colley}
\authornote{Both authors contributed equally to this research.}
\email{m.colley@ucl.ac.uk}
\orcid{0000-0001-5207-5029}
\affiliation{%
  \institution{Institute of Media Informatics, Ulm University}
  \city{Ulm}
  \country{Germany}
}
\affiliation{%
  \institution{UCL Interaction Centre}
  \city{London}
  \country{United Kingdom}
}

\author{Pascal Jansen}
\authornotemark[1]
\email{pascal.jansen@uni-ulm.de}
\orcid{0000-0002-9335-5462}
\affiliation{%
  \institution{Institute of Media Informatics, Ulm University}
  \city{Ulm}
  \country{Germany}
}

\author{Mugdha Keskar}
\email{mugdha.keskar@uni-ulm.de}
\orcid{0009-0008-8450-2759}
\affiliation{%
  \institution{Institute of Media Informatics, Ulm University}
  \city{Ulm}
  \country{Germany}
}

\author{Enrico Rukzio}
\email{enrico.rukzio@uni-ulm.de}
\orcid{0000-0002-4213-2226}
\affiliation{%
  \institution{Institute of Media Informatics, Ulm University}
  \city{Ulm}
  \country{Germany}
}







\renewcommand{\shortauthors}{Colley, Jansen et al.}

\begin{abstract}
The absence of a human operator in automated vehicles (AVs) may require external Human-Machine Interfaces (eHMIs) to facilitate communication with other road users in uncertain scenarios, for example, regarding the right of way.
Given the plethora of adjustable parameters, balancing visual and auditory elements is crucial for effective communication with other road users. With N=37 participants, this study employed multi-objective Bayesian optimization to enhance eHMI designs and improve trust, safety perception, and mental demand. By reporting the Pareto front, we identify optimal design trade-offs. This research contributes to the ongoing standardization efforts of eHMIs, supporting broader adoption.

\end{abstract}



\begin{teaserfigure}
    \includegraphics[width=\linewidth]{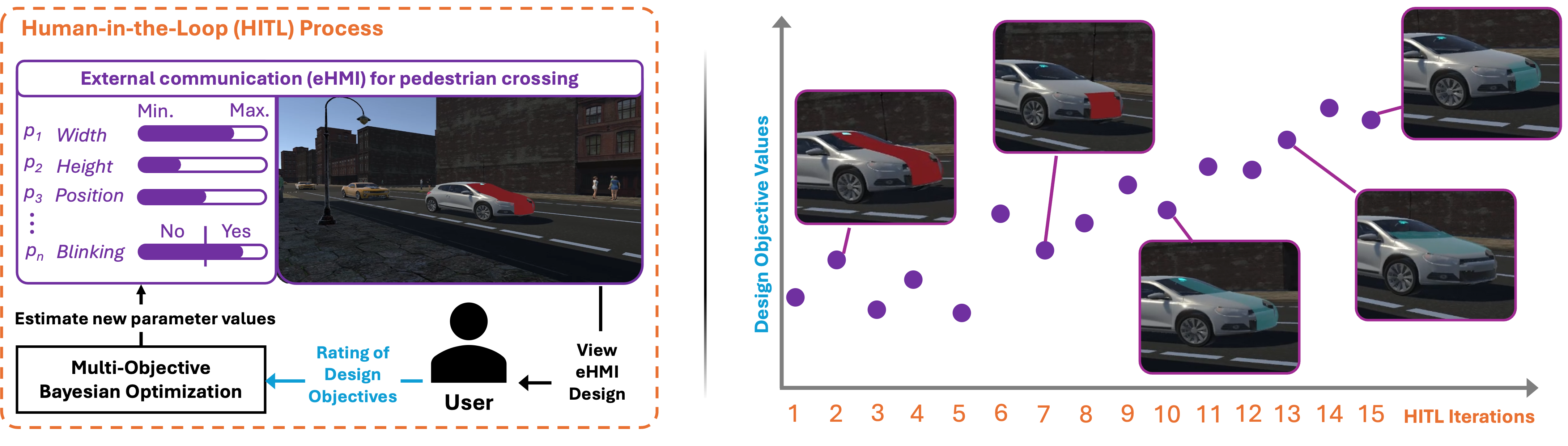}
  \caption{Human-in-the-loop (HITL) MOBO of eHMI design in a pedestrian crossing scenario. The AV communicates its intention to stop and allow pedestrians to cross the street. The goal is to improve user ratings on multiple design objectives, such as trust, safety, acceptance, and aesthetics, while minimizing mental demand. MOBO iteratively uses design parameter values ($p_1$ to $p_n$) and user ratings of the design objectives to suggest optimized designs for each iteration.}
  \label{fig:teaser}
  \Description{Human-in-the-loop (HITL) multi-objective Bayesian optimization (MOBO) of external Human-Machine Interface (eHMI) design in a pedestrian crossing scenario. The automated vehicle communicates its intention to stop and allow pedestrians to cross the street. The goal is to improve user ratings on multiple design objectives, such as trust, safety, acceptance, and aesthetics, while minimizing mental demand. MOBO iteratively uses design parameter values and user ratings of the design objectives to suggest optimized designs for each iteration.}
\end{teaserfigure}


\maketitle

\section{Introduction}
Automated vehicles (AVs) are poised to significantly alter traffic dynamics~\cite{fagnant2015preparing} and interactions within traffic environments~\cite{colley2020adesign, dey2020taming, locken2019should}. Communication tools become essential in the absence of a human operator to communicate with other road users in uncertain situations, such as determining the right of way. These tools are commonly referred to as external Human-Machine Interfaces (eHMIs). Various modalities were proposed for eHMIs, including displays on grilles~\cite{RefWorks:doc:5cf7ad8de4b06bba938e0112} and windshields~\cite{dey2020distancedependent}, LED strips~\cite{RefWorks:doc:5cf7ad8de4b06bba938e0112, RefWorks:doc:5cf8aa47e4b006bc06a90b0a}, movement patterns~\cite{zimmermann2017first}, and projections~\cite{ackermann2019experimental, nguyen2019designing}. External devices like smartphones~\cite{hollander2020smombies} and enhanced infrastructure were considered~\cite{RefWorks:doc:5cd92775e4b0487541989799}. Additionally, various eHMI concepts and designs have been explored, such as text-based designs~\cite{chang2018video} that convey vehicle status~\cite{faas2020longitudinal} or intentions~\cite{dietrich2020automated}. Research generally found positive effects of these eHMIs on trust, clarity, hedonic qualities, and pedestrian crossing behaviors.

Previous work evaluated a limited subset of possible eHMI designs. However, given the numerous design parameters, it is crucial to have a broader picture of the optimal eHMI. In the realm of user interfaces (UIs) for AV passengers, \citet{normark2015design} enabled passengers to manually customize the size, location, and color of icons on the dashboard, center stack, and Head-Up Display (HUD). By tailoring these interfaces to individual preferences, the user experience can be significantly enhanced, leading to increased perceived safety and trust.

Instead of relying on manual adjustments to different UI design aspects, Multi-Objective Bayesian Optimization (MOBO) offers a method to identify optimal design parameters iteratively (see \autoref{fig:teaser}). MOBO has been successfully applied across various domains to address design optimization challenges~\cite{chan2022bo, liao2023interaction, chandramouli2023mobopersonalize, koyama2022boassistant, kadner2021adaptifont}. MOBO predicts which design changes will most effectively meet desired objectives, such as enhancing passengers' trust. It manages multiple objectives by finding the best balance, represented by the Pareto front, ensuring the most effective trade-offs in UI design~\cite{marler_survey_2004}.

In consideration of these factors, this study is guided by the following research question (RQ):
\begin{quote}
\researchq{RQ1}{What are the characteristics of an optimized eHMI for AVs that, among other objectives, enhance pedestrians' trust and perceived safety?}
\end{quote}

To explore this RQ, we conducted a user study with N=37 participants. We applied MOBO to optimize the eHMI's design (see \autoref{fig:teaser}), focusing on the parameters: color, blink frequency, size, positioning, and loudness of the auditory component. These parameters were optimized based on the objectives: trust in automation, understanding, mental demand, perceived safety, acceptance, aesthetics, and the duration until starting to cross.

To explore design preferences across gender, we compared the parameters and objectives based on the participants' gender. Previous work~\cite{o2018gender} already showed that gender impacts crossing decisions. Therefore, \citet{colley2019better} recommended also evaluating this factor for eHMI studies.
Therefore, the second RQ was:
\begin{quote}
\researchq{RQ2}{How do the parameters and objectives differ for gender?}
\end{quote}

In the virtual reality (VR) study, N=37 participants had to cross a two-lane road with traffic in both directions. The traffic was mixed, consisting of AVs and manually driven vehicles. On the near side of the road, AVs communicated their stopping using the eHMI design determined by the MOBO in each iteration.

The comparison of eHMI design parameters between female and male participants revealed no significant differences. However, qualitatively comparing the resulting parameter values on the Pareto front suggests that certain parameter ranges---such as cyan color and a 3Hz flashing animation---could serve as starting points for all users and allow focusing on optimization of other parameters to increase efficiency. Unlike earlier LED strip-based approaches, using the entire front of the AV as an eHMI, combined with a high auditory volume, supports accessibility and multimodality. These findings offer a practical baseline to personalize and refine eHMIs for diverse users.
Participants consistently rated the eHMI highly, reinforcing the need for explicit eHMIs, contrasting with studies questioning their necessity.

\noindent\highlight[purple]{Contribution Statements~\cite{Wobbrock.2016}}
\begin{itemize}[noitemsep, leftmargin=*]
    \item \textbf{Artifact or System} We developed a VR, Unity-based simulation of a pedestrian crossing a non-signalized street, designed to optimize the eHMI through MOBO iteratively, allowing the identification of optimal design parameters based on user feedback across six objectives.
    
    \item \textbf{Empirical study that tells us about how people use a system.} We conducted a between-subjects study (N=37) to investigate the impact of gender on user experience and UI design for eHMIs. Our findings show no significant differences between male and female participants. The MOBO process led to very high objective scores. 
\end{itemize}

\section{Related Work}

We base our work on research on pedestrian-vehicle interactions, focusing on crossing behaviors in traffic and the role and influence of eHMIs in AVs on traffic interactions. Additionally, it presents BO to optimize eHMIs.

\subsection{External Communication of Automated Vehicles}

Current traffic interactions frequently rely on gestures and eye contact to resolve ambiguities~\cite{rasouli2017understanding}. Although explicit communication is rarely required~\cite{lee2021road}, eHMIs were proposed to facilitate interactions between AVs and vulnerable road users such as pedestrians and cyclists~\cite{hollander2021taxonomy}. Previous research has categorized external communication strategies by modality, message type, and vehicle location~\cite{colley2020adesign, colley2020design, dey2020taming}. \citet{colley2020adesign} identified eight types of messages: Instruction, Command, Advisory, Answer, Historical, Predictive, Question, and Affective. Communication can occur at various locations on the vehicle, through personal devices, or via infrastructure like sidewalks, with key interaction points such as the windshield or bumper being particularly important~\cite{dey2019gaze}. Most work and the ISO technical report~\cite{iso23049} recommend communicating the intention of the AV instead of giving advice or commands such as ``Go''. Therefore, our optimization focused on transmitting this information.

Effective eHMI deployment includes considering the communication relationship (ranging from one-to-one to many-to-many), ambient noise levels, and road user (e.g., pedestrian, cyclist)~\cite{colley2020adesign}. Research examined eHMI effectiveness across different groups, including children\cite{deb2019comparison, RefWorks:doc:5cf7c9e4e4b03d2faef34312}, individuals with vision, mobility, or cognitive impairments~\cite{colley2020towards, asha2021co, 10.1145/3546717}, and general pedestrians~\cite{ackermans2020effects, dey2018interface, locken2019should, Dey2024multimodal}, and cyclists~\cite{hou2020autonomous}. Generally, eHMIs have shown positive results. For instance, \citet{dey2020distancedependent} demonstrated that distance-dependent information could significantly improve pedestrians' understanding of AV intentions and willingness to cross safely. \citet{colley2020towards} found that visually impaired individuals preferred clear, speech-based communications over other forms. To address accessibility challenges, our optimization also includes speech output.

However, challenges remain, such as ensuring children accurately interpret and use eHMIs~\cite{deb2019comparison}, concerns about overtrust~\cite{hollander2019overtrust}, scalability~\cite{colley2020scalability, colley2023scalability, 10.1145/3610977.3637478}, and exploration of the social implications of eHMIs~\cite{sadeghian2020exploration, colley2021investigating, lanzer2020designing, sahin2021workshop}. Furthermore, previous work indicated that instead of defining novel eHMIs, research should focus on consolidating previous approaches~\cite{Dey2024multimodal}. With this optimization of well-known parameter constraints (e.g., position), we contribute to this consolidation.

\subsection{Multi-Objective Bayesian Optimization for Interface Designs}
The design of UIs involves selecting parameter values such as element positioning, color, and transparency to achieve specific design objectives like trust and perceived safety. These objectives can be represented as objective functions, with their values (e.g., trust ratings) being determined by specific combinations of design parameters. Since predicting the best parameter combinations is challenging due to the unknown relationship between parameters and objective functions, this problem can be framed as an optimization task~\cite{brochu2010tutorial}.

Given the impracticality of testing all potential design combinations due to the vast design space, BO offers an efficient approach to modeling the relationship between parameter combinations and objective function values. BO is particularly suitable for optimizing unknown and complex functions with limited prior data~\cite{chan2022bo, borji2013bayesian, liao2023human}. It balances \textit{sampling} of new parameter spaces and the \textit{optimization} of known promising regions, leading to optimal designs with fewer iterations.

BO has been applied successfully in various UI design contexts, such as fine-tuning animation appearances~\cite{brochu2010bayesian}, customizing image aesthetics~\cite{koyama2020sequential}, optimizing font settings~\cite{kadner2021adaptifont}, and improving UI interactions~\cite{dudley2019crowdsourcing}. In Human-Computer Interaction (HCI), obtaining objective function values requires user feedback often, making HITL optimization a valuable method. HITL adds a user-centered feedback loop to BO, allowing iterative refinement of design parameters based on user interactions and ratings~\cite{chiu2020human, koyama2020sequential, zhong2021spacewalker}.

However, the design of an eHMI involves multiple objectives, necessitating the use of MOBO. MOBO optimizes several objectives simultaneously, resulting in a Pareto front representing a range of optimal designs, balancing trade-offs between conflicting objectives~\cite{marler_survey_2004}. MOBO has been applied in diverse areas, including multi-finger text entry~\cite{dridhar2015midair} and personalized explanations for image classifiers~\cite{chandramouli2023mobopersonalize}.

Despite its potential, HITL MOBO was not explored in eHMI design with its unique challenges due to subjective objectives like perceived safety and trust, considering the vulnerability of road users like pedestrians and the dynamic nature of traffic, which our work addresses. 

\section{External Communication Concept}
To simulate our eHMI concept, we used Unity (version 2022.3). We opted for a VR simulation using the HTC VIVE Pro, a common method for immersive investigation of eHMIs (e.g., see \citet{colley2021increasing}). VR is an appropriate choice as testing pedestrian crossing scenarios while manipulating the eHMI design can endanger road users in the real world.

\subsection{Scenario}
The pedestrian scenario starts at a sidewalk. The participant needs to cross a two-lane road. Road users in this scenario are manually driven vehicles, AVs, and the participant. On the lane closer to the participant, AVs and manually driven vehicles drive at a 50:50 rate. Only manually driven vehicles are in the far lane. 

After 18s, the next AV on the near lane will stop and let the participant cross. This duration lets the participant experience AVs and manually driven vehicles, representing an externally valid scenario.
The manually driven vehicles on the far lane do not stop.

\begin{figure*}[ht]
    \includegraphics[width=\linewidth]{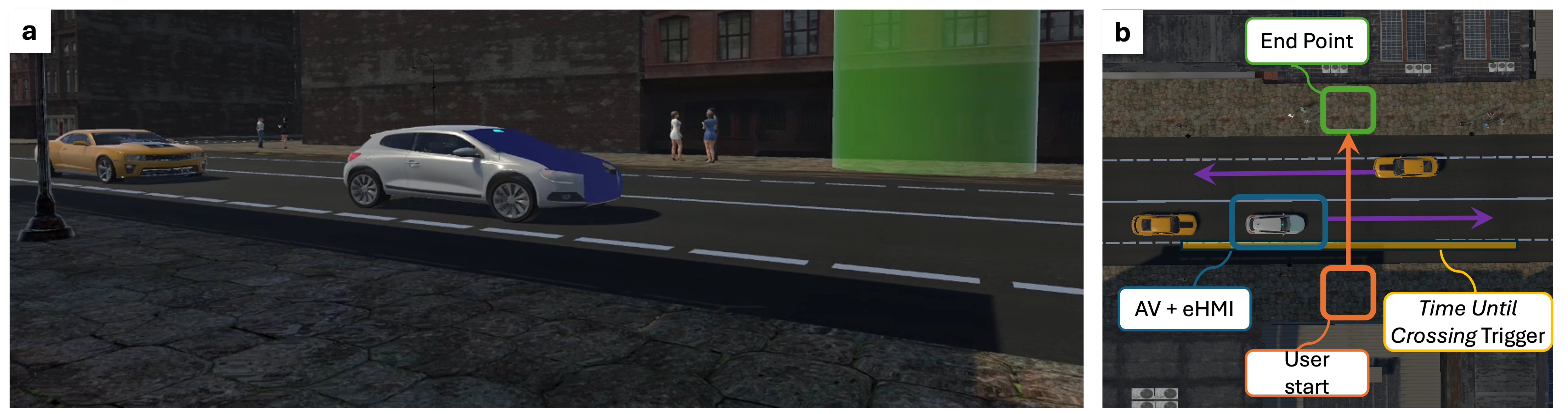}
    \caption{The Unity implementation of the crossing scenario from the participant's starting point of view \textbf{(a)} and from above \textbf{(b)}.}
    \label{fig:pedestrian}
    \Description{This figure has two subfigures. On the left, the participant's point of view is shown. Two vehicles are shown. On the opposite side, a green cone shows where the participant should walk. On the right, the same scene is shown from the top.}
\end{figure*}

\subsection{Generating External Communication}
We defined an eHMI that employs a visual display on the hood, windshield, and grille inspired by eHMIs that use LED light panels (e.g., see \cite{RefWorks:doc:5cf7ad8de4b06bba938e0112, RefWorks:doc:5cf8aa47e4b006bc06a90b0a}). This area is visible from a pedestrian's perspective when driving towards the pedestrian. The area covers the AV's full width and height but excludes headlights and front lights.

\begin{figure*}[ht]
    \includegraphics[width=\linewidth]{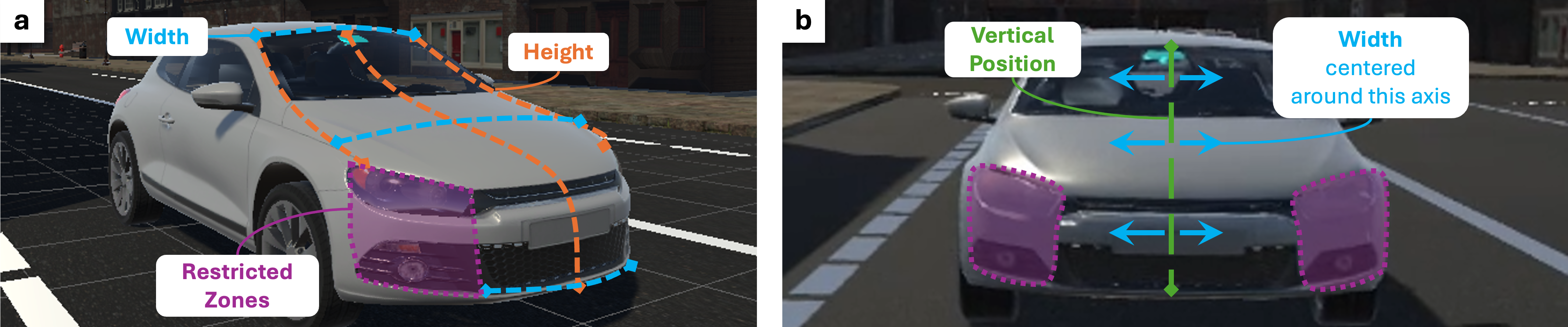}
    \caption{The white AV without an eHMI. A cyan light signal near the rearview mirror indicates that it is an AV. In~\textbf{(a)}, dashed lines show the available \textit{width} and \textit{height} for the eHMI mesh generation, including its lower/upper and left/right bounds. \textbf{(b)} Examples of possible mesh positions in allowed regions. The mesh (i.e., its \textit{width}) is always centered horizontally along the green dashed axis to ensure visibility from both sides. The \textit{vertical position} can be set along this axis. The eHMI does not cover the front lights (restricted zones; in purple).}
    \label{fig:av-car}
    \Description{The white AV is based on a VW Scirocco.}
\end{figure*}

The eHMI's size and position can vary horizontally and vertically, such as a narrow strip on the hood or a small display on the windshield's upper edge. However, the \textit{optimal} position and size for pedestrian communication are unclear. A larger display may be visible from a distance but unclear in its message, while a smaller, more specific display (e.g., on the grille) may be less noticeable.

Therefore, we developed a method to generate an eHMI in Unity that can be positioned and resized dynamically across the windshield, hood, and radiator grille. The parameters and bounds of \textit{width} and \textit{height} for this generation are depicted in \autoref{fig:av-car}. We always center the mesh (i.e., the \textit{width}) horizontally on a \textit{vertical position} axis (see \autoref{fig:av-car} b) for better visibility across the AV’s curved surface if standing directly left/right from the AV. For instance, in \autoref{fig:pedestrian} a, the AV (partly) obstructs a small eHMI mesh on the left side of the hood.

\section{Experiment}
To answer \notextrefrq{RQ1} and \notextrefrq{RQ2}, we conducted an experiment using HITL MOBO (see \autoref{fig:teaser}) to optimize eHMI design in a pedestrian crossing scenario.

\subsection{Bayesian Optimization: Parameterizing the eHMI and the Objective Functions}\label{method-bo}
In the following, we describe the eHMI design parameters and outline our HITL MOBO setup, which iteratively adjusts these toward improving \textit{objective function} values across iterations.

\begin{figure*}[ht]
    \includegraphics[width=\linewidth]{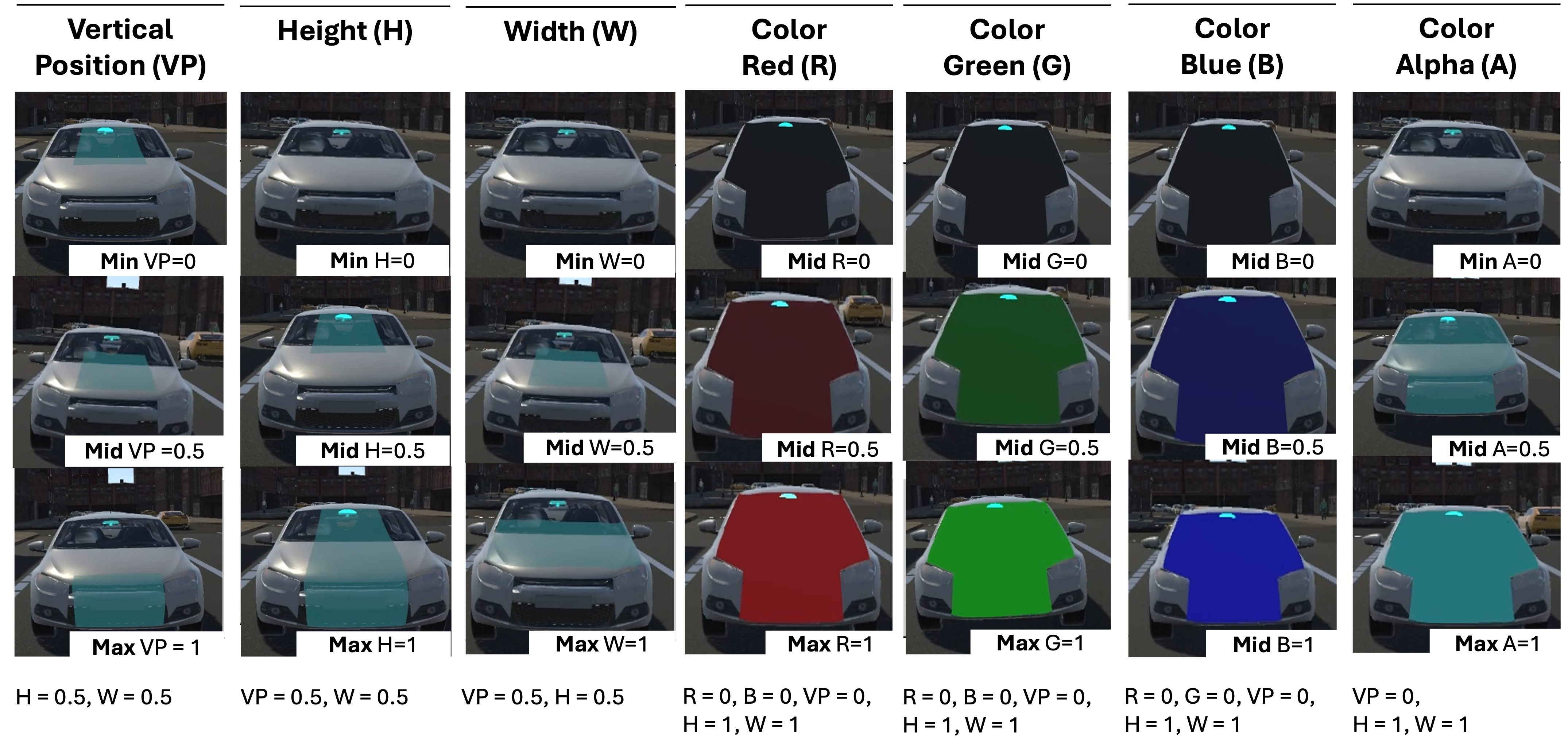}
    \caption{eHMI design parameter value ranges. At the bottom of each column, the default values for the other parameters are shown. Cyan is used as the default color to demonstrate other parameter values.}
    \label{fig:eHMI_examples}
    \Description{Visual representation of the design parameters.}
\end{figure*}

\subsubsection{Design Parameters}\label{method-design-params}
eHMI design parameters were derived from the respective publications~\cite{colley2023scalability, dey2020color, Dey2024multimodal, colley2020adesign, colley2020towards}.
While RGB coloring of eHMIs should avoid unintended meaning (e.g., orange being a warning signal), we included it as a parameter to answer our RQ of the optimal eHMI. 
The alpha level was chosen to allow for the absence of an eHMI (i.e., if the alpha is 0.1, the eHMI is not visible). 
Some work has evaluated blink frequency~\cite{dey2020color}, but the evidence is minimal. The formula for blink frequency is: 
\begin{equation}
\text{Blink frequency (Hz)}  = \frac{1}{\frac{1}{\text{blinkFrequencyMOBO}} \times 0.25}
\label{eq:blink_frequency}
\end{equation}
For example, a value of \textit{0.8} equals 3.2 Hz.
Finally, the size and position of the visual eHMI have to be determined. We allow any given rectangular shape along the outline of the front of the AV. 
Regarding auditory design, we opted for the textual, auditory message ``Stopping'' and varied its loudness. A minimal loudness means no auditory component. 
All design parameters ($p_1$ to $p_{9}$) are summarized in \autoref{tab:design_param}.

Other design elements, such as text-based messages, symbols, projections~\cite{colley2023scalability}, eye-metaphors~\cite{gui2023field}, biology-inspired designs~\cite{oudshoorn2021bio}, or hands~\cite{gui2024shrinkable, mahadevan2018communicating} are possible. However, recent works converged toward an LED stripe, and we assume that manufacturers will focus on such an aesthetic, easy-to-implement, and versatile solution. 

\begin{table*}[ht!]
\scriptsize
\caption{The 9 design parameters for the eHMI design, with ranges. All design parameters are modeled continuously. Example visualizations of parameter values are shown in \autoref{fig:eHMI_examples}.}
\label{tab:design_param}
\resizebox{\textwidth}{!}{%
\begin{tabular}{@{}llll@{}}
\toprule
\textbf{Design Parameter }                  & \textbf{Description}                                  & \textbf{Reference}                 & \textbf{Range}                  \\ \midrule
$p_1$: R       & \textbf{R}ed color channel.   & \cite{colley2020design}  & [0, 1] \\
$p_2$: G       & \textbf{G}reen color channel.   & \cite{colley2020design}  & [0, 1] \\
$p_3$: B       & \textbf{B}lue color channel.   & \cite{colley2020design}  & [0, 1] \\
$p_4$: eHMI Alpha, $\alpha$       & Alpha color channel value.   & \cite{colley2020design}  & [0, 1] \\ \hdashline

$p_5$: Blink frequency & Blink frequency when activated. Maximum: 4Hz.   & \cite{dey2020color}  & [0, 1]             \\

$p_6$: Width, $W$        & Width. Maximum: entire width of the AV. & \cite{colley2020design}   & [0, 1] \\
$p_7$: Height, $H$   &  Height. Maximum: entire height of the AV. & 
 \cite{colley2020design}   & [0, 1]             \\ 
$p_8$: Vertical Position, $VP$        & Position along the center AV axis (viewed from the front). & \cite{colley2020design}   & [0, 1] \\\hdashline

$p_{9}$: Auditory Message Loudness, $l$      & Loudness of auditory message ``Stopping''.       & \cite{colley2020towards}                & [0, 1]\\ \bottomrule
\end{tabular}%
}
\end{table*}

\subsubsection{Objective Functions}\label{method-objectives}

An objective function $f$ maps a specific eHMI design $x$ to a metric the optimizer aims to maximize. 
We focus on maximizing six subjective metrics: \textit{perceived safety}, \textit{trust}, \textit{predictability}, \textit{usefulness}, \textit{satisfaction}, and \textit{visual appeal}. Conversely, we aim to minimize \textit{mental demand} as our sole subjective metric, along with one objective metric: \textit{time to start crossing}.

Based on previous eHMI work~\cite{colley2023scalability}, we employed the following questionnaires after every optimization run in the HITL process:
We assessed \textbf{mental demand} via the mental demand subscale of the raw NASA-TLX~\cite{hart1988development} on a 20-point scale (``How much mental and perceptual activity was required? Was the task easy or demanding, simple or complex?''; 1=\textit{Very Low} to 20=\textit{Very High}; lower is better).
Regarding predictability and trust, we used the subscales \textit{Predictability/Understandability} (\textit{Predictability}) and \textit{Trust} of the \textit{Trust in Automation} questionnaire by \citet{korber2018theoretical}.
\textbf{Predictability} is determined via agreement on four statements (two direct: ``The system state was always clear to me.'', ``I was able to understand why things happened.''; two inverse: ``The system reacts unpredictably.'', ``It's difficult to identify what the system will do next.'') using 5-point Likert scales (1=\textit{Strongly disagree} to 5=\textit{Strongly agree}).
\textbf{Trust} is measured via agreement on the same 5-point Likert scale on two statements (``I trust the system.'' and ``I can rely on the system.''; both times, higher is better).
Participants rated their perceived \textbf{safety} using four 7-point semantic differentials from -3 (anxious/agitated/unsafe/timid) to +3 (relaxed/calm/safe/confident; higher is better)~\cite{faas2020longitudinal}.
Finally, we added three single items.
Two were defined with the van der Laan acceptance scale~\cite{van1997simple} in mind (``I find the visualizations of the automated vehicle \textbf{useful}'', ``I find the visualizations of the automated vehicle \textbf{satisfying}''). These were combined into a single ``acceptance'' objective.
We also adapted the question regarding \textbf{visual appeal} from \citet{colley2023uam} (``I found the visualizations visually appealing''; on a 5-point Likert scale).
\textit{Time to start crossing} is the duration after walking very close to the road since the simulation started. We used a Unity collider to mark the spot where the participant was counted as on the road. (see \autoref{fig:pedestrian} b). 

Normalization into the $[-1, 1]$ range is required because the subjective metrics values have ranges based on 20-, 5-, or 7-point Likert scales.
After transformation, the \textit{mental demand} and \textit{time to start crossing} objectives are to be maximized (a higher value means less load and earlier crossing).

\subsubsection{Hyperparameter Setup for Bayesian Optimization}\label{method-hyperparam-setup}
We used \texttt{BoTorch}~\cite{balandat2020botorch} version 0.11.3 with a multi-output Gaussian Process and applied \texttt{qEHVI} as the acquisition function.
This function represents the expected hypervolume increase, where we set $q=1$ (see \citet{chan2022bo}) to ensure that after each iteration, a batch of size one is selected for evaluation.
The optimization process started with a \textit{sampling} phase of five iterations, during which we employed Sobol sampling \cite{sobol1967} to generate initial eHMI designs. Sobol sampling systematically divides the design space into evenly distributed regions and picks a representative design from each, ensuring broad coverage. At this early stage, the MOBO algorithm has no prior user-specific data, so it needs these initial ratings to build a first understanding of the design space from each individual’s perspective. To prevent bias from different starting conditions, we used the same set of five sampled designs for every participant. After gathering initial ratings, the MOBO used a 15-iteration \textit{optimization} phase, where the optimizer iteratively balanced "exploitation" (refining promising known designs) and "exploration" (probing new regions) for each user. To approximate the acquisition function, we used 2024 restart candidates and 512 Monte Carlo samples (see \citet{chan2022bo}).

\subsubsection{Optimization Stopping Criterion}\label{method-stop-criterion}
In internal tests, we found convergence to an optimal rating of objectives was reached relatively quickly.
Therefore, we added a stopping criterion checked after every measurement: Was the perfect rating for \textbf{every} subjective metric (i.e., the \textbf{highest} rating for trust, predictability, safety, visual appeal, usefulness, satisfaction, and the \textbf{lowest} rating for mental demand; see Section~\ref{method-objectives}) given for the \textbf{last} round? Participants could otherwise not opt out of optimization.

The stopping criterion \textbf{never} applied.

\subsection{Additional Measurements}\label{sec:measurements}
Besides objectives, we logged position with 50Hz, number of collisions with cars, time on the sidewalk and on the street, and total duration. 

Finally, participants assessed the communication (intention to stop, intention not to stop, and timer) regarding necessity and reasonability on individual 7-point Likert scales and gave open feedback.

 \textit{User Expectation Conformity}: "The final design matches my imagination.", 
 \textit{Satisfaction}: "I'm pleased with the final design.", 
 \textit{Confidence}: "I believe the design is optimal for me.", 
 \textit{Agency}: "I felt in control of the design process."  
 \textit{Ownership}: "I feel the final design is mine." Regarding 
 \textit{Interactivity}, participants also provided feedback on desired design control levels ("... Consider aspects where you desired more or less control over the design.")

\subsection{Procedure}

First, participants were introduced to the study procedure and VR scene. They then signed informed consent. The introduction to the setting is in Appendix~\ref{app:introduction}.

\begin{figure*}[ht!]
    \includegraphics[width=\linewidth]{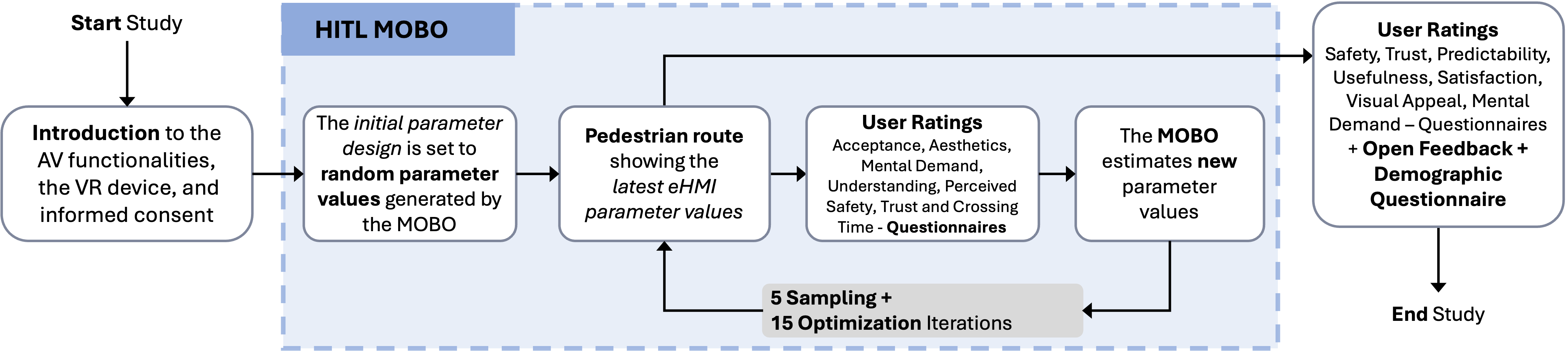}
    \caption{Study procedure using HITL MOBO for eHMI design. There were 20 (5 sampling and 15 optimization) iterations.}
    \label{fig:study-procedure}
    \Description{The study procedure is described in the text with 5 sampling and 15 optimization rounds.}
\end{figure*}

They then signed informed consent and could adjust the headset. 
In each trial, participants crossed the same street with the same setting to guarantee no external effects altering the perception of the eHMI. An AV would stop after $\approx$25 seconds. With crossing the street, each trial took $\approx$1min. 
After each trial, participants answered the questionnaires described in \autoref{sec:measurements} in the VR scene. The study took $\approx$55min. Participants received 10€. The study was conducted in German and English. 

The experimental procedure followed the guidelines of our university's ethics committee and adhered to regulations regarding handling sensitive and private data, anonymization, compensation, and risk aversion. Compliant with our university‘s local regulations, no additional formal ethics approval was required.

\section{Results}

\subsection{Data Analysis}

The goal of MOBO is to identify the Pareto front, which contains all Pareto optimal points in the design space. Each point represents a design that balances trade-offs between conflicting objectives~\cite{marler_survey_2004}. Using the \textit{EMOA} R package~\cite{emoa}, we determined Pareto optimal values for each participant and focused our analysis solely on these efficient designs. 
The female group yielded 76 Pareto designs, and the male group 90.
R 4.4.2 and RStudio 2024.09.1 were employed. All packages were up-to-date in December 2024.

\subsection{Participants}

37 participants (Mean age = 25.6, SD = 3.5, range: [19, 34]; Gender: 45.9\% women, 51.4\% men, 2.70\% non-binary;) took part in the study, recruited locally.
35 participants are college students, two are working, indicating that high school is their highest degree. 
On 5-point Likert scales (\textit{1=Strongly Disagree} --- \textit{5=Strongly Agree}), participants showed interest in AVs  (\m{4.51}, \sd{.77}),  were positive whether AVs would ease their lives (\m{4.40}, \sd{.80}), and were skeptical about whether they become reality by 2034 (\m{4.30}, \sd{1.02}).
For gender comparison, we only used data from female and male participants due to the low number of participants with non-binary gender.

\subsection{Questionnaire Ratings}

\begin{figure*}[ht!]
\centering
\small
    \begin{subfigure}[c]{0.135\linewidth}
        \includegraphics[width=\linewidth]{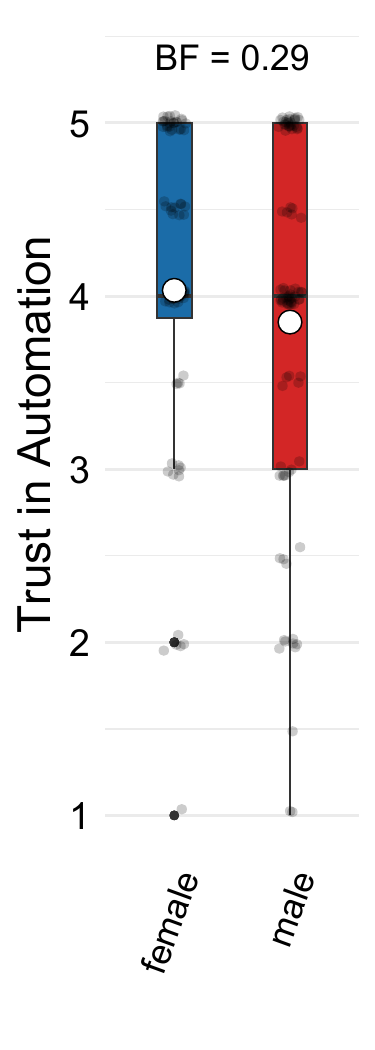}
        \caption{Trust\newline}\label{fig:trust}
        \Description{Trust in Automation}
    \end{subfigure}
    \begin{subfigure}[c]{0.135\linewidth}
        \includegraphics[width=\linewidth]{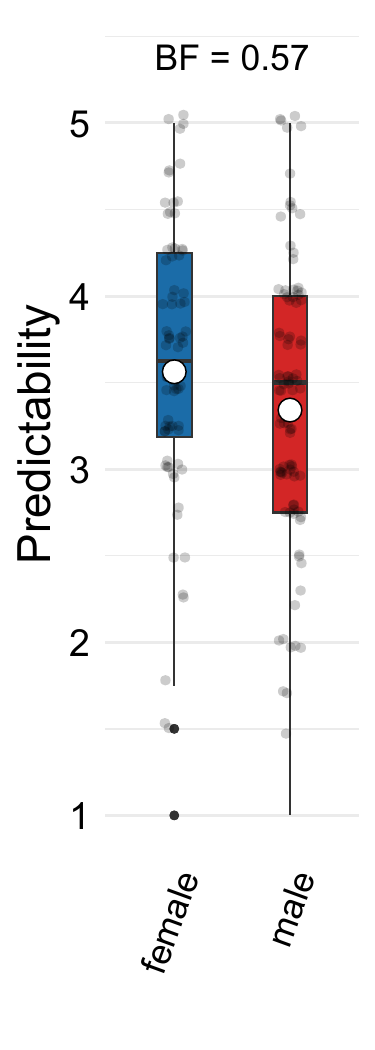}
        \caption{Predictability\newline}\label{fig:understanding}
        \Description{Predictability}
    \end{subfigure}
    \begin{subfigure}[c]{0.135\linewidth}
        \includegraphics[width=\linewidth]{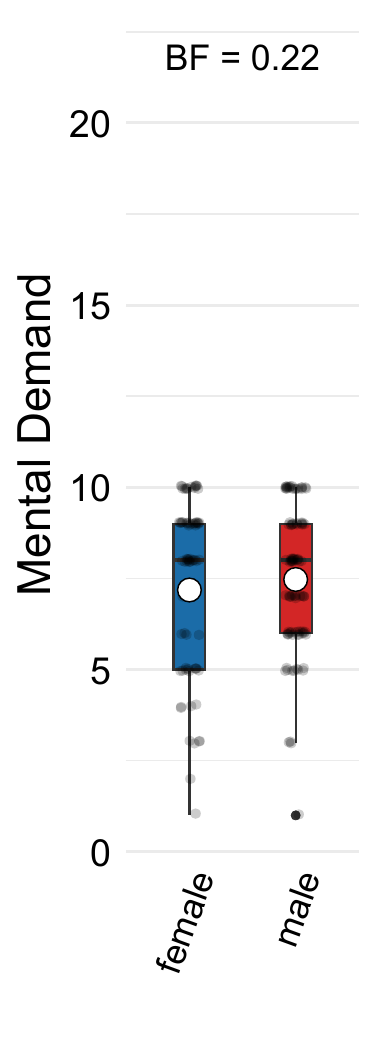}
        \caption{Mental Demand\newline}\label{fig:mental_demand}
        \Description{Mental Demand}
    \end{subfigure}
    \begin{subfigure}[c]{0.135\linewidth}
        \includegraphics[width=\linewidth]{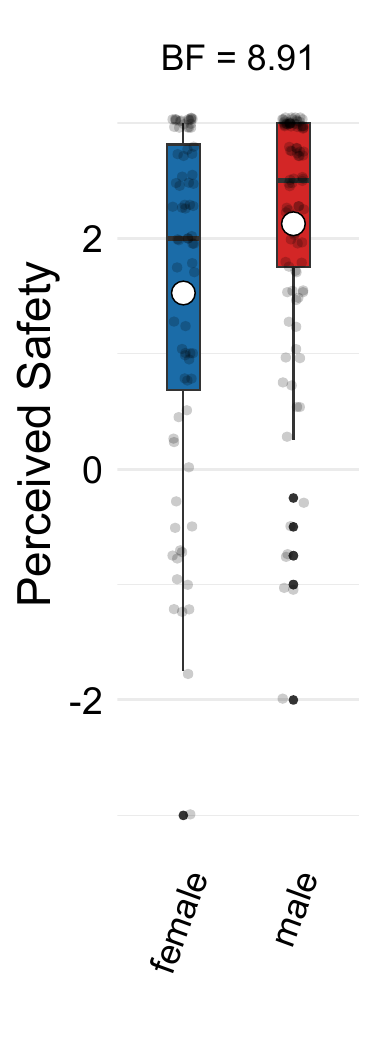}
        \caption{Perceived Safety}\label{fig:perceivedsafety}
        \Description{Perceived Safety}
    \end{subfigure}
    \begin{subfigure}[c]{0.135\linewidth}
        \includegraphics[width=\linewidth]{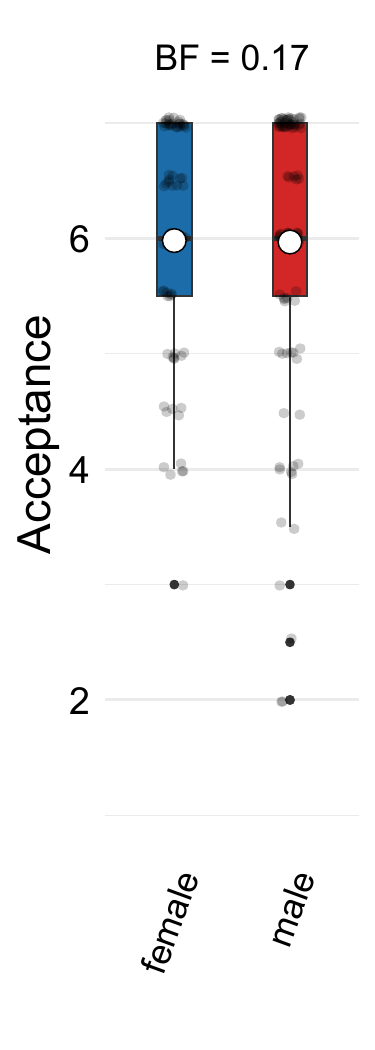}
        \caption{Acceptance\newline}\label{fig:acceptance}
         \Description{Acceptance}
    \end{subfigure}
    \begin{subfigure}[c]{0.135\linewidth}
        \includegraphics[width=\linewidth]{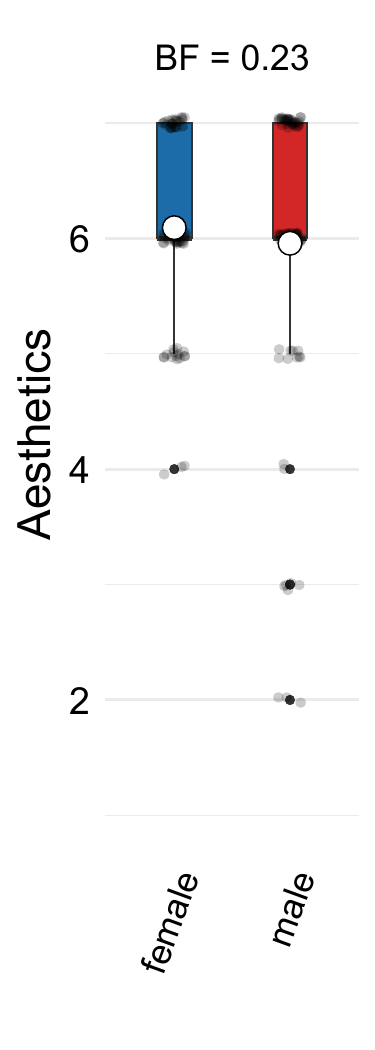}
        \caption{Aesthetics\newline}\label{fig:Aesthetics}
        \Description{Aesthetics}
    \end{subfigure}
        \begin{subfigure}[c]{0.135\linewidth}
        \includegraphics[width=\linewidth]{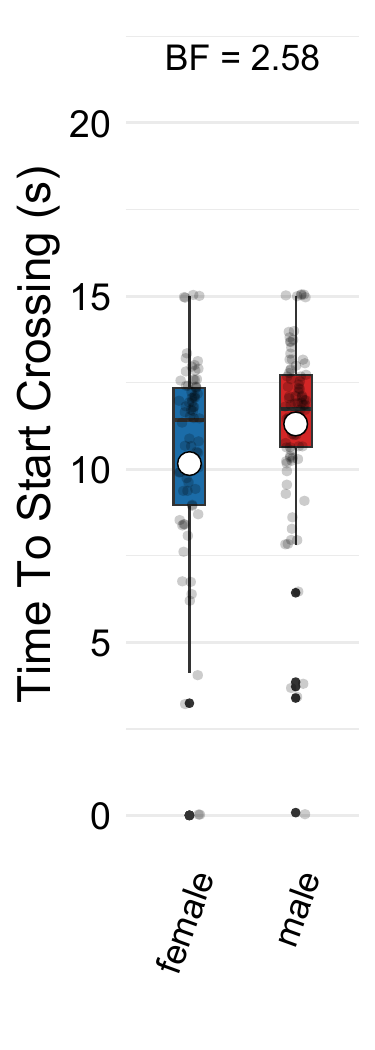}
        \caption{Time To Start Crossing (s)}\label{fig:TimeToStartCrossing}
        \Description{Time To Start Crossing (s)}
    \end{subfigure}

   \caption{Rating for the subjective questionnaires comparing \textit{female} and \textit{male} of Pareto-optimal values. The Bayes factor shows trends towards equality (<1) and difference (>1).}~\label{fig:questionnaire_data}
   \Description{This figure shows the boxplots for comparing both groups' questionnaire ratings (objectives). The figure consists of seven box plots (a-g), each representing a different aspect of user experience measured in the study: Trust in Automation, Predictability, Mental Demand, Perceived Safety, Acceptance, Aesthetics, and Time to start crossing.}
\end{figure*}

We analyzed the mean ratings from the questionnaires of all participants whose design parameters were on the Pareto front. 
\autoref{fig:questionnaire_data} shows a comparison overview. Additionally, we show the value progression in \autoref{fig:runs_acc}, \autoref{fig:runs_aes}, \autoref{fig:runs_cog}, \autoref{fig:runs_pred}, \autoref{fig:runs_ps}, \autoref{fig:runs_trust}, and \autoref{fig:runs_time}, showing that the approach could optimize for the objective values over the iterations. In particular, perceived safety was increased, while mental demand and time to cross decreased. The increase in trust, predictability, and acceptance was weaker. Aesthetics remained roughly constant over the iterations.

\paragraph{Trust in Automation}
The Bayesian analysis of Trust resulted in a $BF = 0.37$ ± 0.00\%, suggesting inconclusive evidence for \textbf{no} difference between females (M=4.03, SD=1.05) and males (M=3.81, SD=1.12; see \autoref{fig:trust}).

\paragraph{Predictability}
For Predictability, the analysis yielded a $BF = 0.91$ ± 0.00\%, providing inconclusive evidence for \textbf{no} differences between the two groups. Females rated Predictability higher (M=3.56, SD=0.88) than males (M=3.30, SD=0.86; see \autoref{fig:understanding}).

\paragraph{Mental Demand} Mental Demand showed a $BF = 0.23$ ± 0.02\%, indicating moderate evidence for equality between the groups. Hence, female ratings (M=7.18, SD=2.65) than male ones (M=7.43, SD=2.08; see \autoref{fig:mental_demand}) for the designs were equal.

\paragraph{Perceived Safety}
Perceived Safety showed a $BF = 6.00$ ± 0.00\%, with moderate evidence favoring difference between the groups. The designs were rated lower for females (M=1.53, SD=1.50) than males (M=2.10, SD=1.14; \autoref{fig:perceivedsafety}).

\paragraph{Acceptance} Acceptance had a $BF = 0.17$ ± 0.05\%, suggesting moderate evidence in favor of \textbf{no} difference between females (M=5.98, SD=1.00) and males (M=5.93, SD=1.25; see \autoref{fig:acceptance}).

\paragraph{Aesthetics}
The Bayesian analysis yielded a $BF = 0.27$ ± 0.04\% for the Aesthetics measure, indicating moderate evidence for \textbf{no} difference. The designs yielded a higher rating for females (M=6.09, SD=0.84) than males (M=5.92, SD=1.27; see \autoref{fig:Aesthetics}).

\paragraph{Time To Start Crossing}
The Bayesian analysis yielded a $BF = 3.50$ ± 0.01\% for Time to start crossing, indicating moderate evidence favoring a difference. The designs yielded a faster start for females (M=10.2, SD=3.55) than males (M=11.4, SD=2.52; see \autoref{fig:TimeToStartCrossing}).

\subsection{Pareto Front Parameter Set}
\begin{figure*}[ht!]
    \centering
    \small
    \includegraphics[width=0.9999\linewidth]{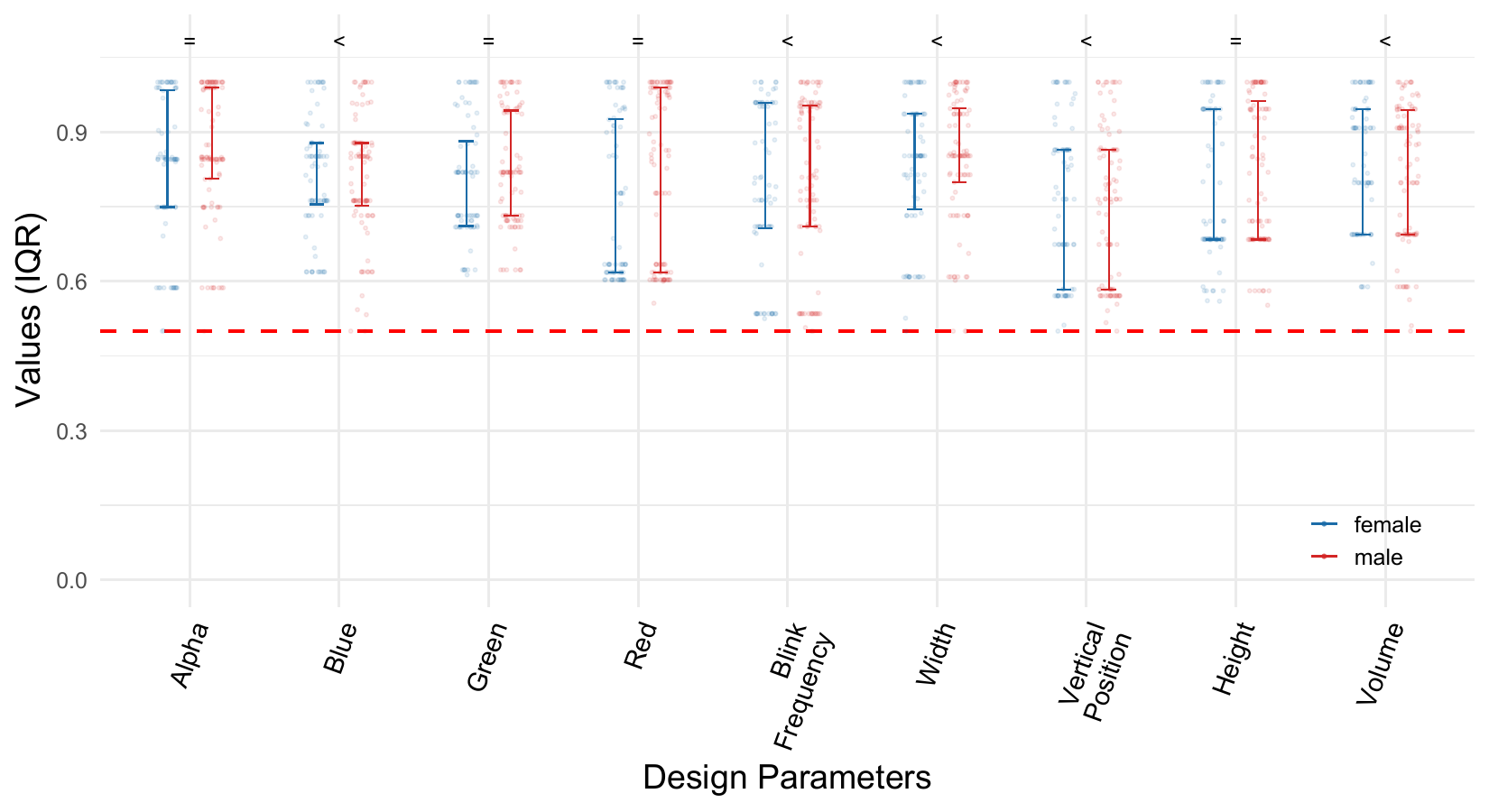}
    \caption{Comparison of the design parameters for both groups, using a Bayesian t-test. The annotations are as defined by \citet{lee2013bayesian}: "$<<<$" for extreme evidence for equality (BF < 0.01), "$<<$" for strong or very strong evidence for equality (BF < 0.1), "$<$" for moderate evidence for equality (BF < 0.3), "$=$" for inconclusive (also called anecdotal) evidence (BF between 0.3 and 3), "$>$" for moderate evidence for difference (BF > 3), "$>>$" for strong or very strong evidence for difference (BF > 10), and "$>>>$" for extreme evidence for difference (BF > 100). The dotted line is a reference line at 0.5}\label{fig:design_parameters}
    \Description{The figure depicts the 9 design parameters to be optimized for the eHMI. A horizontal group of pictures represents each parameter's values from 0, 0.5, and 1.}
\end{figure*}

The results for each design parameter are detailed in \autoref{tab:bf_parameters}. Interestingly, we found no evidence for difference for \textbf{any} parameter. \autoref{fig:design_parameters} shows this visually.

\begin{table*}[ht!]
\centering
\small
\caption{Results of Bayesian Analysis for Each Design Parameter including the IQR ranges}
\begin{tabular}{lllll}
\hline
\textbf{Design Parameter} & \textbf{BF (± \%)} & \textbf{Female IQR} & \textbf{Male IQR} & \textbf{Evidence} \\
\hline
{Alpha} & 0.71 ± 0.02\% & 0.75-0.98 & 0.81-0.99 & inconclusive evid. for equality \\
\textit{Blue} & \textit{0.17 ± 0.05\%} & \textit{0.75-0.88} & \textit{0.75-0.88} & \textit{moderate evid. for equality} \\
{Green} & {0.39 ± 0.03\%} & {0.71-0.88} & {0.73-0.94} & inconclusive evid. for equality \\
{Red} & {1.40 ± 0.01\%} & {0.62-0.93} & {0.62-0.99} & inconclusive evid. for difference \\ \hdashline
\textit{Blink frequency} & \textit{0.19 ± 0.05\%} & \textit{0.71-0.96} & \textit{0.71-0.95} & \textit{moderate evid. for equality} \\
\textit{Width} & \textit{0.22 ± 0.05\%} & \textit{0.74-0.94} & \textit{0.80-0.95} & \textit{moderate evid. for equality} \\
\textit{Vertical position} & \textit{0.19 ± 0.05\%} & \textit{0.58-0.86} & \textit{0.58-0.86} & \textit{moderate evid. for equality} \\
{Height} & 0.37 ± 0.03\% & 0.68-0.95 & 0.68-0.96 & inconclusive evid. for equality \\ \hdashline
\textit{Volume} & \textit{0.21 ± 0.05\%} & \textit{0.69-0.95} & \textit{0.69-0.94} & \textit{moderate evid. for equality} \\
\hline
\end{tabular}
\label{tab:bf_parameters}
\end{table*}

\begin{figure*}[ht!]
\centering
    \includegraphics[width=\linewidth]{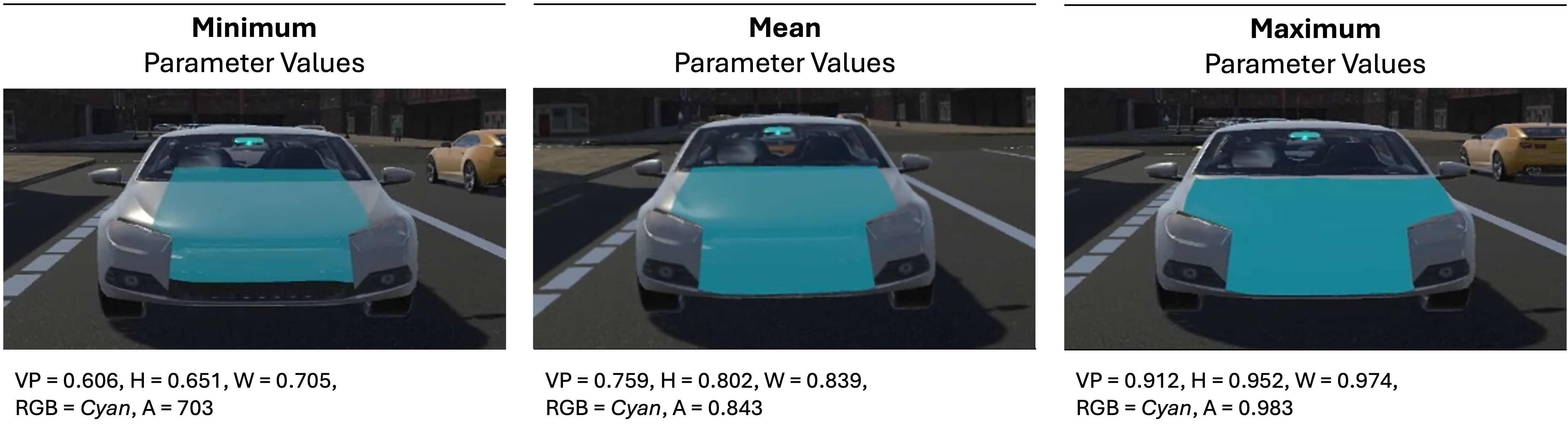}
   \caption{Visualizations of the resulting design parameter value ranges on the \textbf{Pareto front}. Cyan was used as a default color (RGB) to demonstrate the non-color-related values: vertical position (VP), height (H), width (W), and alpha (A). Minimum/Maximum shows the lowest/highest values across participants where the design received Pareto-optimal ratings.}
   \label{fig:eHMI-results-example}
    \Description{Three subfigures show the minimum, mean, and maximum design parameter values for the Pareto front. The color was fixed to cyan.}
\end{figure*}

\autoref{fig:eHMI-results-example} visualizes the final mean parameter set on the Pareto front in the scene. We use the commonly used color cyan~\cite{dey2020color} as a reference. The values between minimum, mean, and maximum do not vary widely.

\begin{figure*}[ht!]
\centering
    \includegraphics[width=0.95\linewidth]{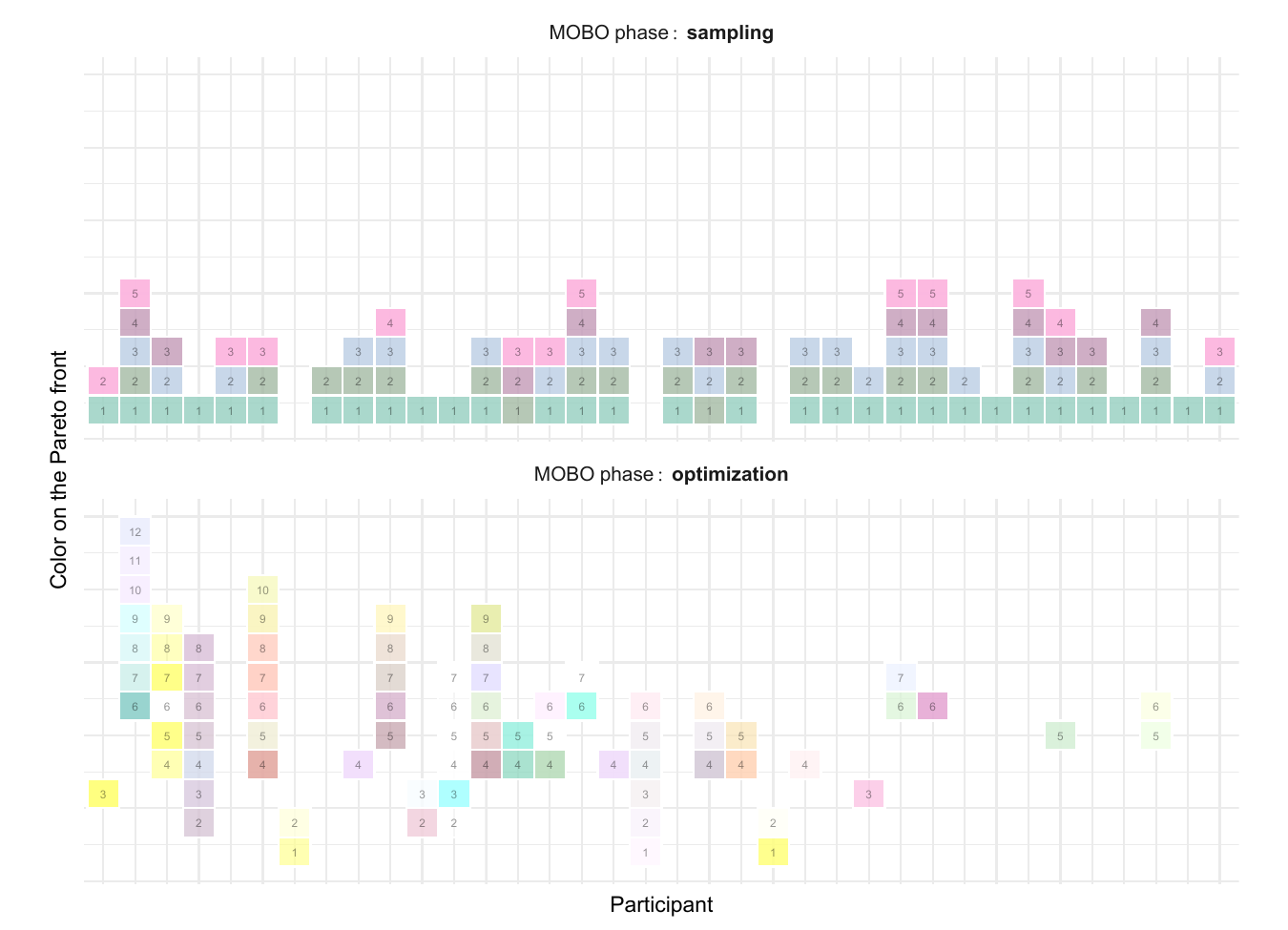}
   \caption{Color distribution for the \textbf{Pareto front}, split into sampling (1-5) and optimization (1-15) phases. The number in a box indicates the iteration in which this color was set, counting from the bottom. The columns on the x-axis represent the individual participants. An empty spot in a column means this iteration was \textbf{not} Pareto-true (i.e., dominated by another design).}
   \label{fig:color}
    \Description{Color distribution for the Pareto front per participant and per MOBO phase.}
\end{figure*}

Because averaging RGB values would obscure meaningful differences, we plot all resulting color values on the Pareto-front per participant across the 20 iterations in \autoref{fig:color}. Each column shows a single participant’s Pareto-true color designs over time. Each colored box represents one iteration where a given design was Pareto-optimal (i.e., non-dominated), and empty spaces indicate dominated designs and, thus, not on the Pareto front. While various colors emerged, hues close to cyan were consistently present. This recurrence suggests that ''standard'' cyan hues remain a robust choice even in a process driven by iterative optimization. During sampling (iterations 1–5), in which all users saw the same initial five designs, 34 out of 37 participants had at least one Pareto-optimal color, suggesting that the initial sampling effectively spanned the design space. However, fewer participants achieved Pareto-optimal colors during optimization (iterations 1–15), where MOBO focused on refining promising regions. Some participants even had more than five Pareto-optimal colors in this phase, highlighting individual preferences and indicating that more iterations may be needed to identify optimal colors for certain users reliably.

\subsection{Correlation between the Objectives}

\begin{figure}[ht!]
    \centering
    \small
    \begin{subfigure}[b]{0.49\textwidth}
        \centering
        \includegraphics[width=\textwidth]{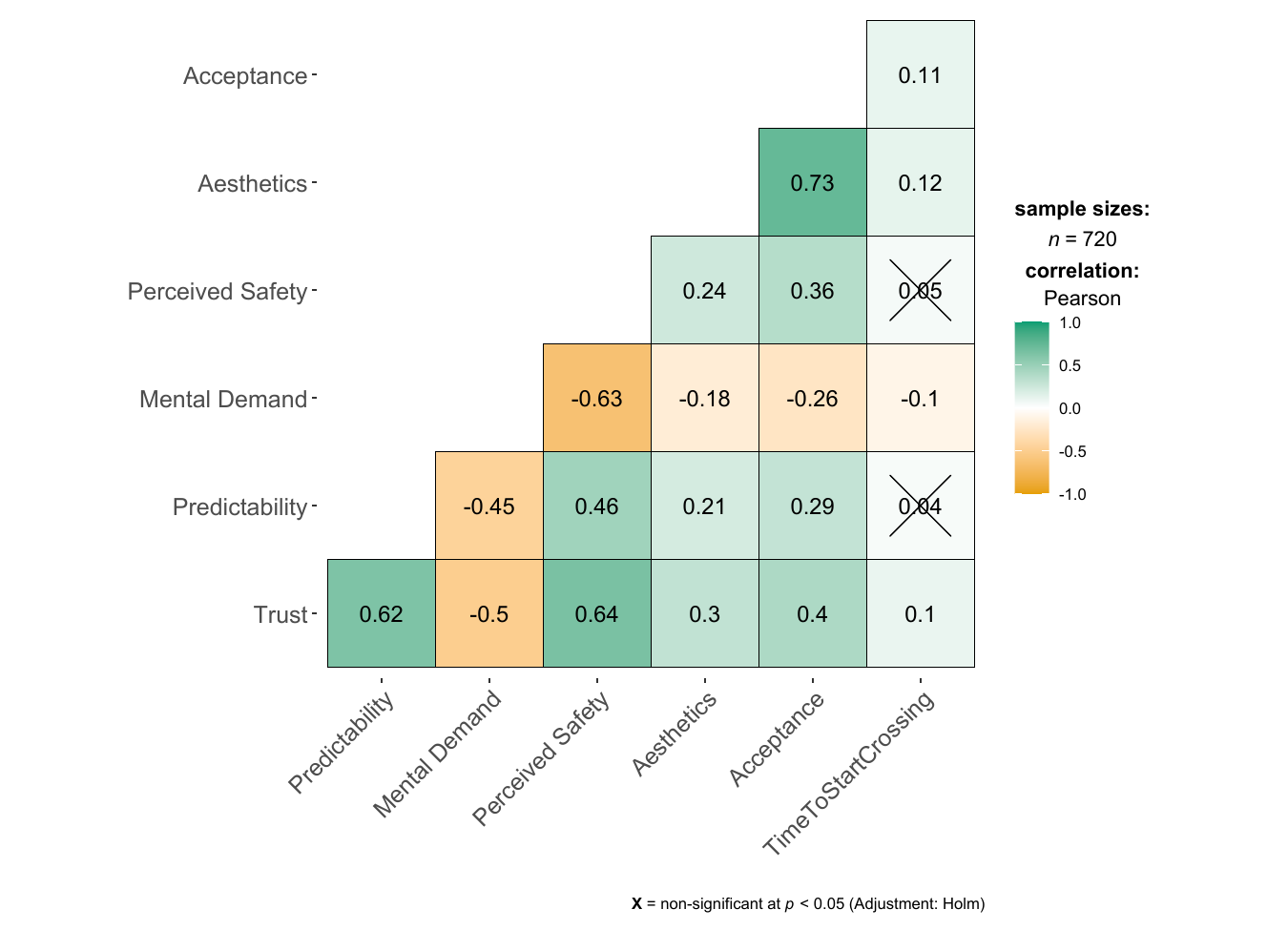}
        \caption{All objectives.}
        \label{fig:correlation_all}
    \end{subfigure}
    \hfill
    \begin{subfigure}[b]{0.49\textwidth}
        \centering
        \includegraphics[width=\textwidth]{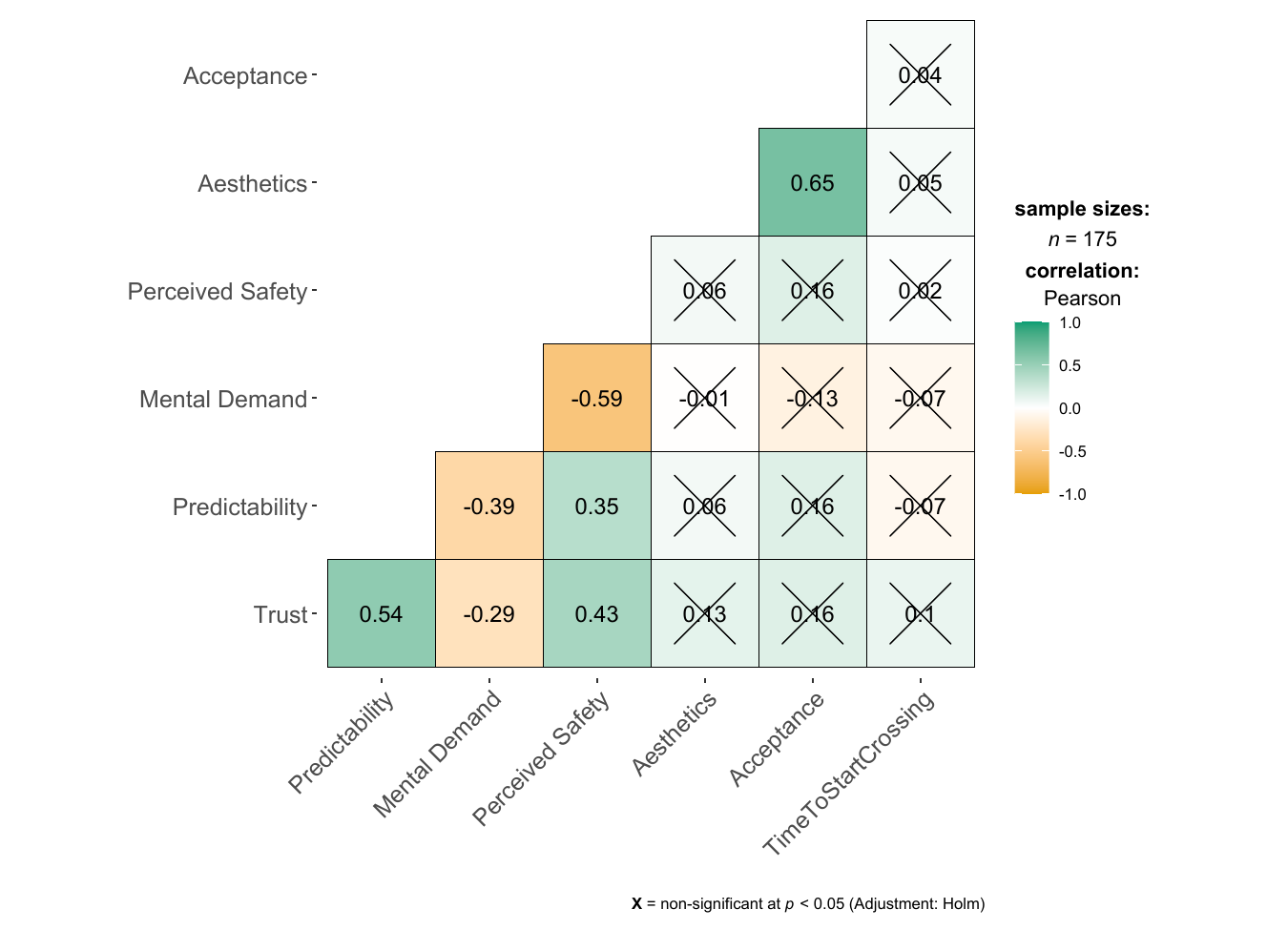}
        \caption{Only Pareto-true objectives.}
        \label{fig:correlation_mobo}
    \end{subfigure}
    \caption{Correlation heatmaps of study objectives. (a) includes all data points ($n=720=20 * 37$, i.e., 20 iterations for 37 participants), and (b) includes only Pareto-true data points. ''$x$'' indicates non-significant at $p<0.05$ (Adjustment: Holm).}
    \label{fig:correlation_combined}
        \Description{This figure is a correlation heatmap, displaying the relationships between the objectives used in the study: Trust, Predictability, Mental Demand, Perceived Safety, Aesthetics, Acceptance, and Time to start crossing. The figure uses a color gradient, with lighter shades representing weaker correlations and darker shades indicating stronger correlations. The correlation values range from -1.0 to 1.0, where positive values indicate a positive correlation and negative values indicate a negative correlation.}
\end{figure}

MOBO identifies optimal design parameters that balance multiple objectives along the Pareto front, ensuring that no single objective can be improved without compromising others~\cite{marler_survey_2004}. To assess whether there are trade-offs between these objectives, we calculated the correlations among all objectives (see \autoref{fig:correlation_combined}). This analysis helps us understand how changes in one objective might influence others.

The results show that almost all correlations were statistically significant when using all data (see \autoref{fig:correlation_all}). In particular, Trust and Predictability had a strong positive correlation ($r=0.65$), as did acceptance and aesthetics ($r=0.71$), meaning improvements in one were associated with improvements in the other. In contrast, the correlations between Mental Demand and the other objectives were negative. 
There were no correlations between Time to start crossing with Predictability and Perceived Safety.

When only taking Pareto front values into account (see \autoref{fig:correlation_mobo}), much fewer significant correlations exist. However, Trust still had a strong positive correlation with Predictability ($r=0.54$) and Perceived Safety ($r=0.43$), for example. Compared to when all data was used, all correlations became less strong and no correlation inverted.


\subsection{Design Satisfaction and Engagement}
Participants evaluated several design process aspects using Likert scales ranging from 1 (Strongly Disagree) to 7 (Strongly Agree). 

Firstly, participants reported that the final design matched their expectations (``The final design matches my expectation.''; \m{5.70}, \sd{1.58}). This suggests that most users felt the outcome aligned with their anticipated expectations. Similarly, satisfaction (``I'm pleased with the final design.'') with the final design was high (\m{6.00}, \sd{1.41}).

Control over the design process (``I felt in control of the design process.'') was another positive aspect (\m{5.81}, \sd{1.35}).  Confidence in the optimality of the design (``I believe the design is optimal for me.'') for their needs was also high (\m{5.97}, \sd{1.30}).

However, ownership of the final design (``I feel the final design is mine.''), while still positive, showed a slightly lower mean of \m{5.30} (\sd{1.61}). 

Overall, the data suggests that participants were generally pleased with the design and felt a strong sense of control and confidence in the process, but the degree of ownership felt towards the final design may warrant further exploration to ensure greater personal investment in the outcomes.

\subsection{Open Feedback}

Participants provided detailed feedback regarding their experience with the design and interaction during the study. Several key themes emerged, including interactivity, general impressions, and areas for improvement.

\textit{Interactivity and Control:}
While many participants felt the design was effective, some expressed uncertainty about their level of control over the process. One participant mentioned, \textit{"I did not realise I was controlling the design process,"} highlighting a disconnect between the design and user awareness of control. Another echoed a similar sentiment, \textit{"I would like more control over how the HMI looked in general. The range of adjustments was too small and easy to miss sometimes."}

\textit{General Impressions:}
Overall, feedback was positive, with several participants praising the design. One stated, 
\textit{"Overall, the study and the design of the eHMI was effective. It was well paced, and organised."} Another appreciated the clarity of the vehicle's communication, saying, \textit{"The announcement of stop/go was useful alongside the concept of colors. It made it easier for pedestrians to understand when to and not to cross the road."} The audio and visual cues were particularly well-received, with a participant noting, \textit{"The sound 'stopping' is really a good reminder for me to cross the road, let me feel safe."}

\textit{Areas for Improvement:}   
Despite positive feedback, several suggestions for improvement were made. One participant mentioned the unpredictability of the vehicle’s stopping behavior: \textit{"Not sure when the automated car decides it wants to stop. The eHMI seems to come out suddenly."} Another recommended more gradual cues, suggesting, \textit{"A gradual light/indicator that the car is going to slow down for me—recognizes and conveys that intent."} Some participants also noted that the interaction could benefit from additional elements, such as \textit{"more interaction elements in the scenario, or building more than one."}

A few participants commented on the unpredictability of the vehicles, stating, \textit{"In the beginning, it was not clear for me where exactly I have to stand and wait for the vehicles to stop."} Others felt that different colors or light effects would enhance communication, with one suggesting, \textit{"The lights were adequate. However, a bright light in the day and night would be better, like a bright green or red color."}


\section{Discussion}
While numerous works evaluated different designs of eHMIs~\cite{colley2020adesign, dey2020color, tran2024exploring}, there is the open question of how an optimal eHMI should be designed. 

We conducted a between-subject study to investigate these RQs with 37 participants. We employed MOBO to iteratively optimize the eHMI designs based on participant ratings of trust, perceived safety, understanding, mental demand, acceptance, aesthetics, and time to start crossing. This method enabled us to compare designs on the Pareto, where enhancing one objective would necessitate trade-offs with others~\cite{marler_survey_2004}. In the following, we discuss our findings regarding the optimal eHMI and the design process. 

\subsection{Optimized Design Choices in External Communication of Automated Vehicles}
The comparison of optimized eHMI design parameters between females and males revealed \textbf{no} evidence for differences.
Regarding \notextrefrq{RQ2}, however, this does not necessarily imply a generally coherent Pareto front for all users.
Gender is only one demographic factor, and its absence of effect in our study does not necessarily mean that a single ''universal'' eHMI design would work equally well for all. User characteristics are complex, influenced by factors such as age, cultural background~\cite{10.1145/3699778}, cognitive abilities, and personal experiences (e.g., see \citet{edelmann2021cross}).

Yet, looking more qualitatively at the final design parameters (see \autoref{fig:design_parameters}), we conclude that the Pareto front incorporates a similar design. 
Therefore, we posit that a certain set of parameters---e.g., cyan hue, flashing animation around 3Hz, and a large illuminated area on the vehicle’s front---might be appropriate for most users, at least as an initial starting design from which personalization is more efficient.
We argue in line with previous work~\cite{colley2023scalability, dey2020color} that cyan is an appropriate color. We modeled the preferred flashing animation~\cite{dey2020color} and found that $\approx$3Hz was optimal, compared to 1Hz used by \citet{dey2020color}. Compared to previous designs that used LED strips~\cite{dey2020color, colley2023scalability, 10.1145/3678506}, we enabled the entire AV front to be an eHMI (also see \citet{schlackl2020ehmi} or \citet{dey2020distancedependent}). 

Interestingly, \autoref{fig:design_parameters} shows that on the Pareto front, the volume of the auditory message was relatively high (approximately 0.8, with no value on the Pareto front below 0.5) both for female and male participants. Moderate evidence for similarity was also found here for female and male participants. This is important for accessibility considerations~\cite{colley2020towards} and also replicates findings that support the positive effect of multimodality (e.g., visual and auditory signals) of eHMIs~\cite{Dey2024multimodal}. This furthermore highlights the possibility of having a "one-fits-all" eHMI~\cite{locken2020wecare} for some design parameters. 

This work also advocates \textbf{for} an eHMI as with (at least) 20 rounds of interaction, an eHMI was still rated highly. While perceived safety most likely also increased due to increased exposure time~\cite{10.1145/3491102.3517571}, this is still a finding supporting explicit eHMIs and is in contrast to work considering explicit eHMIs not necessary~\cite{moore2019case}.

\subsection{Individual Optimization and the Notion of Universality}
General approaches exist to focus on \textit{group level} optimization. For example, one can use the causal tree analysis by \citet{athey2016recursive} and then the derived groups as a homogenous group for optimization. However, it is unclear which variables should be used for the grouping in the context of eHMIs. We have, therefore, used gender as a known characteristic affecting the crossing decision~\cite{colley2019better}. Interestingly, we found \textbf{no} significant differences in the objectives or the design parameters. 
However, this does not mean that genuine individual differences are absent; rather, it indicates that our chosen demographic factor and current sample might not have captured meaningful variability. 

\subsection{Trade-off in the Objectives}\label{sec:trade-off} 
MOBO seeks to identify optimal design parameters that balance multiple objectives along the Pareto front, ensuring that enhancing one objective does not disproportionately compromise another~\cite{marler_survey_2004}. This approach is particularly valuable when trade-offs exist between objectives, such as spatial error versus completion time~\cite{chan2022bo}. However, our analysis uncovered strong correlations among objectives, suggesting minimal conflict between some.

This observation implies potential redundancy within the current set of objectives. Although these were selected based on previous eHMI studies~\cite{colley2020scalability, colley2020towards, dey2020color, dey2020distancedependent}, future research could streamline the objectives by prioritizing those that are most distinct and impactful. For example, trust and predictability or acceptance and aesthetics show strong correlations (see \autoref{fig:correlation_combined}), indicating that selecting one from each pair could simplify future studies. Conversely, mental demand, which exhibits only a weak correlation with Perceived safety when looking at Pareto optimal values (see \autoref{fig:correlation_mobo}), should be preserved as it captures a unique user experience.

Although these are preliminary insights, future research should further investigate and validate the potential for reducing objectives in eHMI studies.

\subsection{Practical Guidelines for eHMI Designs}

Our study's findings provide valuable practical insights for future research on eHMIs. Here are the two key guidelines:

\begin{itemize}
    \item \textbf{Multimodality is Key:} When designing eHMIs, accessibility of traffic is improved when employing multimodality~\cite{colley2020towards} and also wanted by people without disabilities. 

    \item \textbf{Starting Points for Efficient Personalization:} While personalization and continuous optimization remain necessary to account for individual differences (apart from gender), our findings suggest certain parameter ranges as effective starting points. These could enable the optimization to focus on fewer other parameters to yield eHMI designs that quickly align with diverse user needs.
    
\end{itemize}

\subsection{Limitations and Future Work}

We prioritized subjective ratings as main objectives for optimizing the UI design parameters. However, we included one objective value: Time to start crossing. It remains to be discussed whether more objective data, such as objective physiological responses, is useful. Nonetheless, the subjective ratings were collected using validated questionnaires or inspired by related research. While valid, future research should investigate the relationship between these dimensions and more specific subcomponents of trust. Additionally, it is crucial to consider the potential redundancy of some questionnaire items due to high correlations between certain measures (see Section~\ref{sec:trade-off}).

Our analysis provided inconclusive evidence on gender differences, limiting the strength of our claims. 

Future work should expand the age range and incorporate more diverse participant pools. Although we included German and Indian participants, future investigations should broaden cultural contexts~\cite{edelmann2021cross}.
Future work would also benefit from testing the eHMI designs under various environmental conditions, such as daytime versus nighttime or weather scenarios (see \citet{colley2020adesign}), to better assess their effectiveness in real-world applications. Additionally, the scope of the eHMI designs could be expanded by incorporating other design elements, such as text-based messages, symbols, or dynamic animations. Prior work has, however, shown a high potential for LED stripes in eHMI designs due to their easy integration and aesthetically pleasing design. Text, for example, is inaccessible to children or foreigners, and symbols can be unclear. We used the flashing animation as prior work showed a preference for this pattern~\cite{dey2020color}.

\section{Conclusion}
This study explored the eHMI design preferences using the MOBO approach. We conducted a study with N=37 participants in VR. Our findings indicate no differences between female and male participants. While this does not yet claim that a universal design for all users exists, assessing the design parameters on the Pareto front, found that some design parameter value ranges seem feasible for future work to focus on. Therefore, this work supports the standardization efforts necessary in the automotive domain.

\section*{Open Science}
We make the Bayesian optimizer (see \url{https://github.com/Pascal-Jansen/Bayesian-Optimization-for-Unity}), the Unity application upon request, and the collected (anonymized) data available (see \url{https://github.com/M-Colley/ehmi-optimization-chi25-data}). 

\begin{acks}
The authors thank all study participants.
\end{acks}

\bibliographystyle{ACM-Reference-Format}
\bibliography{ehmi-bo.bib}


\begin{thebibliography}{80}


\ifx \showCODEN    \undefined \def \showCODEN     #1{\unskip}     \fi
\ifx \showISBNx    \undefined \def \showISBNx     #1{\unskip}     \fi
\ifx \showISBNxiii \undefined \def \showISBNxiii  #1{\unskip}     \fi
\ifx \showISSN     \undefined \def \showISSN      #1{\unskip}     \fi
\ifx \showLCCN     \undefined \def \showLCCN      #1{\unskip}     \fi
\ifx \shownote     \undefined \def \shownote      #1{#1}          \fi
\ifx \showarticletitle \undefined \def \showarticletitle #1{#1}   \fi
\ifx \showURL      \undefined \def \showURL       {\relax}        \fi
\providecommand\bibfield[2]{#2}
\providecommand\bibinfo[2]{#2}
\providecommand\natexlab[1]{#1}
\providecommand\showeprint[2][]{arXiv:#2}

\bibitem[23049:2018(2018)]%
        {iso23049}
\bibfield{author}{\bibinfo{person}{ISO/TR 23049:2018}.} \bibinfo{year}{2018}\natexlab{}.
\newblock \bibinfo{booktitle}{\emph{Road Vehicles: Ergonomic Aspects of External Visual Communication from Automated Vehicles to Other Road Users.}}
\newblock \bibinfo{type}{Standard}. \bibinfo{institution}{International Organization for Standardization}.
\newblock


\bibitem[Ackermann et~al\mbox{.}(2019)]%
        {ackermann2019experimental}
\bibfield{author}{\bibinfo{person}{Claudia Ackermann}, \bibinfo{person}{Matthias Beggiato}, \bibinfo{person}{Sarah Schubert}, {and} \bibinfo{person}{Josef~F Krems}.} \bibinfo{year}{2019}\natexlab{}.
\newblock \showarticletitle{An experimental study to investigate design and assessment criteria: what is important for communication between pedestrians and automated vehicles?}
\newblock \bibinfo{journal}{\emph{Applied ergonomics}}  \bibinfo{volume}{75} (\bibinfo{year}{2019}), \bibinfo{pages}{272--282}.
\newblock


\bibitem[Ackermans et~al\mbox{.}(2020)]%
        {ackermans2020effects}
\bibfield{author}{\bibinfo{person}{Sander Ackermans}, \bibinfo{person}{Debargha Dey}, \bibinfo{person}{Peter Ruijten}, \bibinfo{person}{Raymond~H. Cuijpers}, {and} \bibinfo{person}{Bastian Pfleging}.} \bibinfo{year}{2020}\natexlab{}.
\newblock \showarticletitle{The Effects of Explicit Intention Communication, Conspicuous Sensors, and Pedestrian Attitude in Interactions with Automated Vehicles}. In \bibinfo{booktitle}{\emph{Proceedings of the 2020 CHI Conference on Human Factors in Computing Systems}} (Honolulu, HI, USA) \emph{(\bibinfo{series}{CHI ’20})}. \bibinfo{publisher}{ACM}, \bibinfo{address}{New York, NY, USA}, \bibinfo{pages}{1–14}.
\newblock
\showISBNx{9781450367080}
\href{https://doi.org/10.1145/3313831.3376197}{doi:\nolinkurl{10.1145/3313831.3376197}}


\bibitem[Asha et~al\mbox{.}(2021)]%
        {asha2021co}
\bibfield{author}{\bibinfo{person}{Ashratuz~Zavin Asha}, \bibinfo{person}{Christopher Smith}, \bibinfo{person}{Georgina Freeman}, \bibinfo{person}{Sean Crump}, \bibinfo{person}{Sowmya Somanath}, \bibinfo{person}{Lora Oehlberg}, {and} \bibinfo{person}{Ehud Sharlin}.} \bibinfo{year}{2021}\natexlab{}.
\newblock \showarticletitle{Co-Designing Interactions between Pedestrians in Wheelchairs and Autonomous Vehicles}. In \bibinfo{booktitle}{\emph{Designing Interactive Systems Conference 2021}} (Virtual Event, USA) \emph{(\bibinfo{series}{DIS '21})}. \bibinfo{publisher}{ACM}, \bibinfo{address}{New York, NY, USA}, \bibinfo{pages}{339–351}.
\newblock
\showISBNx{9781450384766}
\href{https://doi.org/10.1145/3461778.3462068}{doi:\nolinkurl{10.1145/3461778.3462068}}


\bibitem[Athey and Imbens(2016)]%
        {athey2016recursive}
\bibfield{author}{\bibinfo{person}{Susan Athey} {and} \bibinfo{person}{Guido Imbens}.} \bibinfo{year}{2016}\natexlab{}.
\newblock \showarticletitle{Recursive partitioning for heterogeneous causal effects}.
\newblock \bibinfo{journal}{\emph{Proceedings of the National Academy of Sciences}} \bibinfo{volume}{113}, \bibinfo{number}{27} (\bibinfo{year}{2016}), \bibinfo{pages}{7353--7360}.
\newblock


\bibitem[Balandat et~al\mbox{.}(2020)]%
        {balandat2020botorch}
\bibfield{author}{\bibinfo{person}{Maximilian Balandat}, \bibinfo{person}{Brian Karrer}, \bibinfo{person}{Daniel~R. Jiang}, \bibinfo{person}{Samuel Daulton}, \bibinfo{person}{Benjamin Letham}, \bibinfo{person}{Andrew~Gordon Wilson}, {and} \bibinfo{person}{Eytan Bakshy}.} \bibinfo{year}{2020}\natexlab{}.
\newblock \showarticletitle{{BoTorch: A Framework for Efficient Monte-Carlo Bayesian Optimization}}. In \bibinfo{booktitle}{\emph{Advances in Neural Information Processing Systems 33}}.
\newblock
\urldef\tempurl%
\url{http://arxiv.org/abs/1910.06403}
\showURL{%
\tempurl}


\bibitem[Borji and Itti(2013)]%
        {borji2013bayesian}
\bibfield{author}{\bibinfo{person}{Ali Borji} {and} \bibinfo{person}{Laurent Itti}.} \bibinfo{year}{2013}\natexlab{}.
\newblock \showarticletitle{Bayesian optimization explains human active search}.
\newblock \bibinfo{journal}{\emph{Advances in neural information processing systems}}  \bibinfo{volume}{26} (\bibinfo{year}{2013}).
\newblock


\bibitem[Brochu et~al\mbox{.}(2010a)]%
        {brochu2010bayesian}
\bibfield{author}{\bibinfo{person}{Eric Brochu}, \bibinfo{person}{Tyson Brochu}, {and} \bibinfo{person}{Nando De~Freitas}.} \bibinfo{year}{2010}\natexlab{a}.
\newblock \showarticletitle{A Bayesian interactive optimization approach to procedural animation design}. In \bibinfo{booktitle}{\emph{Proceedings of the 2010 ACM SIGGRAPH/Eurographics Symposium on Computer Animation}}. \bibinfo{publisher}{ACM}, \bibinfo{address}{New York, NY, USA}, \bibinfo{pages}{103--112}.
\newblock


\bibitem[Brochu et~al\mbox{.}(2010b)]%
        {brochu2010tutorial}
\bibfield{author}{\bibinfo{person}{Eric Brochu}, \bibinfo{person}{Vlad~M. Cora}, {and} \bibinfo{person}{Nando de Freitas}.} \bibinfo{year}{2010}\natexlab{b}.
\newblock \bibinfo{title}{A {Tutorial} on {Bayesian} {Optimization} of {Expensive} {Cost} {Functions}, with {Application} to {Active} {User} {Modeling} and {Hierarchical} {Reinforcement} {Learning}}.
\newblock
\urldef\tempurl%
\url{http://arxiv.org/abs/1012.2599}
\showURL{%
\tempurl}
\newblock
\shownote{arXiv:1012.2599 [cs]}.


\bibitem[Chan et~al\mbox{.}(2022)]%
        {chan2022bo}
\bibfield{author}{\bibinfo{person}{Liwei Chan}, \bibinfo{person}{Yi-Chi Liao}, \bibinfo{person}{George~B Mo}, \bibinfo{person}{John~J Dudley}, \bibinfo{person}{Chun-Lien Cheng}, \bibinfo{person}{Per~Ola Kristensson}, {and} \bibinfo{person}{Antti Oulasvirta}.} \bibinfo{year}{2022}\natexlab{}.
\newblock \showarticletitle{Investigating Positive and Negative Qualities of Human-in-the-Loop Optimization for Designing Interaction Techniques}. In \bibinfo{booktitle}{\emph{Proceedings of the 2022 CHI Conference on Human Factors in Computing Systems}} (New Orleans, LA, USA) \emph{(\bibinfo{series}{CHI '22})}. \bibinfo{publisher}{ACM}, \bibinfo{address}{New York, NY, USA}, Article \bibinfo{articleno}{112}, \bibinfo{numpages}{14}~pages.
\newblock
\showISBNx{9781450391573}
\href{https://doi.org/10.1145/3491102.3501850}{doi:\nolinkurl{10.1145/3491102.3501850}}


\bibitem[Chandramouli et~al\mbox{.}(2023)]%
        {chandramouli2023mobopersonalize}
\bibfield{author}{\bibinfo{person}{Suyog Chandramouli}, \bibinfo{person}{Yifan Zhu}, {and} \bibinfo{person}{Antti Oulasvirta}.} \bibinfo{year}{2023}\natexlab{}.
\newblock \showarticletitle{Interactive Personalization of Classifiers for Explainability Using Multi-Objective Bayesian Optimization}. In \bibinfo{booktitle}{\emph{Proceedings of the 31st ACM Conference on User Modeling, Adaptation and Personalization}} (Limassol, Cyprus) \emph{(\bibinfo{series}{UMAP '23})}. \bibinfo{publisher}{ACM}, \bibinfo{address}{New York, NY, USA}, \bibinfo{pages}{34–45}.
\newblock
\showISBNx{9781450399326}
\href{https://doi.org/10.1145/3565472.3592956}{doi:\nolinkurl{10.1145/3565472.3592956}}


\bibitem[Chang et~al\mbox{.}(2018)]%
        {chang2018video}
\bibfield{author}{\bibinfo{person}{Chia-Ming Chang}, \bibinfo{person}{Koki Toda}, \bibinfo{person}{Takeo Igarashi}, \bibinfo{person}{Masahiro Miyata}, {and} \bibinfo{person}{Yasuhiro Kobayashi}.} \bibinfo{year}{2018}\natexlab{}.
\newblock \showarticletitle{A Video-Based Study Comparing Communication Modalities between an Autonomous Car and a Pedestrian}. In \bibinfo{booktitle}{\emph{Adjunct Proceedings of the 10th International Conference on Automotive User Interfaces and Interactive Vehicular Applications}} (Toronto, ON, Canada) \emph{(\bibinfo{series}{AutomotiveUI ’18})}. \bibinfo{publisher}{ACM}, \bibinfo{address}{New York, NY, USA}, \bibinfo{pages}{104–109}.
\newblock
\showISBNx{9781450359474}
\href{https://doi.org/10.1145/3239092.3265950}{doi:\nolinkurl{10.1145/3239092.3265950}}


\bibitem[Charisi et~al\mbox{.}(2017)]%
        {RefWorks:doc:5cf7c9e4e4b03d2faef34312}
\bibfield{author}{\bibinfo{person}{Vicky Charisi}, \bibinfo{person}{Azra Habibovic}, \bibinfo{person}{Jonas Andersson}, \bibinfo{person}{Jamy Li}, {and} \bibinfo{person}{Vanessa Evers}.} \bibinfo{year}{2017}\natexlab{}.
\newblock \showarticletitle{Children’s Views on Identification and Intention Communication of Self-Driving Vehicles}. In \bibinfo{booktitle}{\emph{Proceedings of the 2017 Conference on Interaction Design and Children}} (Stanford, California, USA) \emph{(\bibinfo{series}{IDC ’17})}. \bibinfo{publisher}{ACM}, \bibinfo{address}{New York, NY, USA}, \bibinfo{pages}{399–404}.
\newblock
\showISBNx{9781450349215}
\href{https://doi.org/10.1145/3078072.3084300}{doi:\nolinkurl{10.1145/3078072.3084300}}


\bibitem[Chiu et~al\mbox{.}(2020)]%
        {chiu2020human}
\bibfield{author}{\bibinfo{person}{Chia-Hsing Chiu}, \bibinfo{person}{Yuki Koyama}, \bibinfo{person}{Yu-Chi Lai}, \bibinfo{person}{Takeo Igarashi}, {and} \bibinfo{person}{Yonghao Yue}.} \bibinfo{year}{2020}\natexlab{}.
\newblock \showarticletitle{Human-in-the-Loop Differential Subspace Search in High-Dimensional Latent Space}.
\newblock \bibinfo{journal}{\emph{ACM Trans. Graph.}} \bibinfo{volume}{39}, \bibinfo{number}{4}, Article \bibinfo{articleno}{85} (\bibinfo{date}{aug} \bibinfo{year}{2020}), \bibinfo{numpages}{15}~pages.
\newblock
\showISSN{0730-0301}
\href{https://doi.org/10.1145/3386569.3392409}{doi:\nolinkurl{10.1145/3386569.3392409}}


\bibitem[Colley et~al\mbox{.}(2022)]%
        {10.1145/3491102.3517571}
\bibfield{author}{\bibinfo{person}{Mark Colley}, \bibinfo{person}{Elvedin Bajrovic}, {and} \bibinfo{person}{Enrico Rukzio}.} \bibinfo{year}{2022}\natexlab{}.
\newblock \showarticletitle{Effects of Pedestrian Behavior, Time Pressure, and Repeated Exposure on Crossing Decisions in Front of Automated Vehicles Equipped with External Communication}. In \bibinfo{booktitle}{\emph{Proceedings of the 2022 CHI Conference on Human Factors in Computing Systems}} (New Orleans, LA, USA) \emph{(\bibinfo{series}{CHI '22})}. \bibinfo{publisher}{ACM}, \bibinfo{address}{New York, NY, USA}, Article \bibinfo{articleno}{367}, \bibinfo{numpages}{11}~pages.
\newblock
\showISBNx{9781450391573}
\href{https://doi.org/10.1145/3491102.3517571}{doi:\nolinkurl{10.1145/3491102.3517571}}


\bibitem[Colley et~al\mbox{.}(2021a)]%
        {colley2021investigating}
\bibfield{author}{\bibinfo{person}{Mark Colley}, \bibinfo{person}{Jan~Henry Belz}, {and} \bibinfo{person}{Enrico Rukzio}.} \bibinfo{year}{2021}\natexlab{a}.
\newblock \showarticletitle{Investigating the Effects of Feedback Communication of Autonomous Vehicles}. In \bibinfo{booktitle}{\emph{13th International Conference on Automotive User Interfaces and Interactive Vehicular Applications}}. \bibinfo{publisher}{ACM}, \bibinfo{address}{New York, NY, USA}, \bibinfo{pages}{263–273}.
\newblock
\showISBNx{9781450380638}
\urldef\tempurl%
\url{https://doi.org/10.1145/3409118.3475133}
\showURL{%
\tempurl}


\bibitem[Colley et~al\mbox{.}(2023a)]%
        {colley2023scalability}
\bibfield{author}{\bibinfo{person}{Mark Colley}, \bibinfo{person}{Julian Britten}, {and} \bibinfo{person}{Enrico Rukzio}.} \bibinfo{year}{2023}\natexlab{a}.
\newblock \showarticletitle{Scalability in External Communication of Automated Vehicles: Evaluation and Recommendations}.
\newblock \bibinfo{journal}{\emph{Proceedings of the ACM on Interactive, Mobile, Wearable and Ubiquitous Technologies}} \bibinfo{volume}{7}, \bibinfo{number}{2} (\bibinfo{year}{2023}), \bibinfo{pages}{1--26}.
\newblock


\bibitem[Colley et~al\mbox{.}(2024a)]%
        {10.1145/3610977.3637478}
\bibfield{author}{\bibinfo{person}{Mark Colley}, \bibinfo{person}{Julian Czymmeck}, \bibinfo{person}{Mustafa K\"{u}c\"{u}kkocak}, \bibinfo{person}{Pascal Jansen}, {and} \bibinfo{person}{Enrico Rukzio}.} \bibinfo{year}{2024}\natexlab{a}.
\newblock \showarticletitle{PedSUMO: Simulacra of Automated Vehicle-Pedestrian Interaction Using SUMO To Study Large-Scale Effects}. In \bibinfo{booktitle}{\emph{Proceedings of the 2024 ACM/IEEE International Conference on Human-Robot Interaction}} (Boulder, CO, USA) \emph{(\bibinfo{series}{HRI '24})}. \bibinfo{publisher}{ACM}, \bibinfo{address}{New York, NY, USA}, \bibinfo{pages}{890–895}.
\newblock
\showISBNx{9798400703225}
\href{https://doi.org/10.1145/3610977.3637478}{doi:\nolinkurl{10.1145/3610977.3637478}}


\bibitem[Colley et~al\mbox{.}(2024b)]%
        {10.1145/3699778}
\bibfield{author}{\bibinfo{person}{Mark Colley}, \bibinfo{person}{Daniel Kornm\"{u}ller}, \bibinfo{person}{Debargha Dey}, \bibinfo{person}{Wendy Ju}, {and} \bibinfo{person}{Enrico Rukzio}.} \bibinfo{year}{2024}\natexlab{b}.
\newblock \showarticletitle{Longitudinal Effects of External Communication of Automated Vehicles in the USA and Germany: A Comparative Study in Virtual Reality and Via a Browser}.
\newblock \bibinfo{journal}{\emph{Proc. ACM Interact. Mob. Wearable Ubiquitous Technol.}} \bibinfo{volume}{8}, \bibinfo{number}{4}, Article \bibinfo{articleno}{176} (\bibinfo{date}{Nov.} \bibinfo{year}{2024}), \bibinfo{numpages}{33}~pages.
\newblock
\href{https://doi.org/10.1145/3699778}{doi:\nolinkurl{10.1145/3699778}}


\bibitem[Colley et~al\mbox{.}(2021b)]%
        {colley2021increasing}
\bibfield{author}{\bibinfo{person}{Mark Colley}, \bibinfo{person}{Surong Li}, {and} \bibinfo{person}{Enrico Rukzio}.} \bibinfo{year}{2021}\natexlab{b}.
\newblock \showarticletitle{Increasing Pedestrian Safety Using External Communication of Autonomous Vehicles for Signalling Hazards}. In \bibinfo{booktitle}{\emph{Proceedings of the 23rd International Conference on Mobile Human-Computer Interaction}}. \bibinfo{publisher}{ACM}, \bibinfo{address}{New York, NY, USA}, Article \bibinfo{articleno}{20}, \bibinfo{numpages}{10}~pages.
\newblock
\showISBNx{9781450383288}
\urldef\tempurl%
\url{https://doi.org/10.1145/3447526.3472024}
\showURL{%
\tempurl}


\bibitem[Colley et~al\mbox{.}(2023b)]%
        {colley2023uam}
\bibfield{author}{\bibinfo{person}{Mark Colley}, \bibinfo{person}{Luca-Maxim Meinhardt}, \bibinfo{person}{Alexander Fassbender}, \bibinfo{person}{Michael Rietzler}, {and} \bibinfo{person}{Enrico Rukzio}.} \bibinfo{year}{2023}\natexlab{b}.
\newblock \showarticletitle{Come Fly With Me: Investigating the Effects of Path Visualizations in Automated Urban Air Mobility}.
\newblock \bibinfo{journal}{\emph{Proc. ACM Interact. Mob. Wearable Ubiquitous Technol.}} \bibinfo{volume}{7}, \bibinfo{number}{2}, Article \bibinfo{articleno}{52} (\bibinfo{date}{jun} \bibinfo{year}{2023}), \bibinfo{numpages}{23}~pages.
\newblock
\href{https://doi.org/10.1145/3596249}{doi:\nolinkurl{10.1145/3596249}}


\bibitem[Colley and Rukzio(2020a)]%
        {colley2020adesign}
\bibfield{author}{\bibinfo{person}{Mark Colley} {and} \bibinfo{person}{Enrico Rukzio}.} \bibinfo{year}{2020}\natexlab{a}.
\newblock \showarticletitle{A Design Space for External Communication of Autonomous Vehicles}. In \bibinfo{booktitle}{\emph{Proceedings of the 12th International Conference on Automotive User Interfaces and Interactive Vehicular Applications}} \emph{(\bibinfo{series}{AutomotiveUI ’20})}. \bibinfo{publisher}{ACM}, \bibinfo{address}{New York, NY, USA}.
\newblock
\href{https://doi.org/10.1145/3409120.3410646}{doi:\nolinkurl{10.1145/3409120.3410646}}


\bibitem[Colley and Rukzio(2020b)]%
        {colley2020design}
\bibfield{author}{\bibinfo{person}{Mark Colley} {and} \bibinfo{person}{Enrico Rukzio}.} \bibinfo{year}{2020}\natexlab{b}.
\newblock \showarticletitle{Towards a Design Space for External Communication of Autonomous Vehicles}. In \bibinfo{booktitle}{\emph{Extended Abstracts of the 2020 CHI Conference on Human Factors in Computing Systems}} (Honolulu, Hawaii USA) \emph{(\bibinfo{series}{CHI ’20})}. \bibinfo{publisher}{ACM}, \bibinfo{address}{New York, NY, USA}.
\newblock
\href{https://doi.org/10.1145/3334480.3382844}{doi:\nolinkurl{10.1145/3334480.3382844}}


\bibitem[Colley et~al\mbox{.}(2020b)]%
        {colley2020towards}
\bibfield{author}{\bibinfo{person}{Mark Colley}, \bibinfo{person}{Marcel Walch}, \bibinfo{person}{Jan Gugenheimer}, \bibinfo{person}{Ali Askari}, {and} \bibinfo{person}{Enrico Rukzio}.} \bibinfo{year}{2020}\natexlab{b}.
\newblock \showarticletitle{Towards Inclusive External Communication of Autonomous Vehicles for Pedestrians with Vision Impairments}. In \bibinfo{booktitle}{\emph{Proceedings of the 2020 CHI Conference on Human Factors in Computing Systems}} (Honolulu, HI, USA) \emph{(\bibinfo{series}{CHI ’20})}. \bibinfo{publisher}{ACM}, \bibinfo{address}{New York, NY, USA}, \bibinfo{pages}{1–14}.
\newblock
\showISBNx{9781450367080}
\href{https://doi.org/10.1145/3313831.3376472}{doi:\nolinkurl{10.1145/3313831.3376472}}


\bibitem[Colley et~al\mbox{.}(2019)]%
        {colley2019better}
\bibfield{author}{\bibinfo{person}{Mark Colley}, \bibinfo{person}{Marcel Walch}, {and} \bibinfo{person}{Enrico Rukzio}.} \bibinfo{year}{2019}\natexlab{}.
\newblock \showarticletitle{For a Better (Simulated) World: Considerations for VR in External Communication Research}. In \bibinfo{booktitle}{\emph{Proceedings of the 11th International Conference on Automotive User Interfaces and Interactive Vehicular Applications: Adjunct Proceedings}} (Utrecht, Netherlands) \emph{(\bibinfo{series}{AutomotiveUI ’19})}. \bibinfo{publisher}{ACM}, \bibinfo{address}{New York, NY, USA}, \bibinfo{pages}{442–449}.
\newblock
\showISBNx{9781450369206}
\href{https://doi.org/10.1145/3349263.3351523}{doi:\nolinkurl{10.1145/3349263.3351523}}


\bibitem[Colley et~al\mbox{.}(2020a)]%
        {colley2020scalability}
\bibfield{author}{\bibinfo{person}{Mark Colley}, \bibinfo{person}{Marcel Walch}, {and} \bibinfo{person}{Enrico Rukzio}.} \bibinfo{year}{2020}\natexlab{a}.
\newblock \showarticletitle{Unveiling the Lack of Scalability in Research on External Communication of Autonomous Vehicles}. In \bibinfo{booktitle}{\emph{Extended Abstracts of the 2020 CHI Conference on Human Factors in Computing Systems}} (Honolulu, Hawaii USA) \emph{(\bibinfo{series}{CHI ’20})}. \bibinfo{publisher}{ACM}, \bibinfo{address}{New York, NY, USA}.
\newblock
\href{https://doi.org/10.1145/3334480.3382865}{doi:\nolinkurl{10.1145/3334480.3382865}}


\bibitem[Deb et~al\mbox{.}(2020)]%
        {deb2019comparison}
\bibfield{author}{\bibinfo{person}{Shuchisnigdha Deb}, \bibinfo{person}{Daniel~W. Carruth}, \bibinfo{person}{Muztaba Fuad}, \bibinfo{person}{Laura~M. Stanley}, {and} \bibinfo{person}{Darren Frey}.} \bibinfo{year}{2020}\natexlab{}.
\newblock \showarticletitle{Comparison of Child and Adult Pedestrian Perspectives of External Features on Autonomous Vehicles Using Virtual Reality Experiment}. In \bibinfo{booktitle}{\emph{Advances in Human Factors of Transportation}}, \bibfield{editor}{\bibinfo{person}{Neville Stanton}} (Ed.). \bibinfo{publisher}{Springer International Publishing}, \bibinfo{address}{Cham}, \bibinfo{pages}{145--156}.
\newblock
\showISBNx{978-3-030-20503-4}


\bibitem[Dey et~al\mbox{.}(2020a)]%
        {dey2020taming}
\bibfield{author}{\bibinfo{person}{Debargha Dey}, \bibinfo{person}{Azra Habibovic}, \bibinfo{person}{Andreas L{\"o}cken}, \bibinfo{person}{Philipp Wintersberger}, \bibinfo{person}{Bastian Pfleging}, \bibinfo{person}{Andreas Riener}, \bibinfo{person}{Marieke Martens}, {and} \bibinfo{person}{Jacques Terken}.} \bibinfo{year}{2020}\natexlab{a}.
\newblock \showarticletitle{Taming the eHMI jungle: A classification taxonomy to guide, compare, and assess the design principles of automated vehicles' external human-machine interfaces}.
\newblock \bibinfo{journal}{\emph{Transportation Research Interdisciplinary Perspectives}}  \bibinfo{volume}{7} (\bibinfo{year}{2020}), \bibinfo{pages}{100174}.
\newblock


\bibitem[Dey et~al\mbox{.}(2020b)]%
        {dey2020color}
\bibfield{author}{\bibinfo{person}{Debargha Dey}, \bibinfo{person}{Azra Habibovic}, \bibinfo{person}{Bastian Pfleging}, \bibinfo{person}{Marieke Martens}, {and} \bibinfo{person}{Jacques Terken}.} \bibinfo{year}{2020}\natexlab{b}.
\newblock \showarticletitle{Color and Animation Preferences for a Light Band {{eHMI}} in Interactions between Automated Vehicles and Pedestrians}. In \bibinfo{booktitle}{\emph{Proceedings of the 2020 {{CHI Conference}} on {{Human Factors}} in {{Computing Systems}}}}. \bibinfo{publisher}{ACM}, \bibinfo{address}{New York, NY, USA}, \bibinfo{pages}{1--13}.
\newblock


\bibitem[Dey et~al\mbox{.}(2020c)]%
        {dey2020distancedependent}
\bibfield{author}{\bibinfo{person}{Debargha Dey}, \bibinfo{person}{Kai Holl\"{a}nder}, \bibinfo{person}{Melanie Berger}, \bibinfo{person}{Berry Eggen}, \bibinfo{person}{Marieke Martens}, \bibinfo{person}{Bastian Pfleging}, {and} \bibinfo{person}{Jacques Terken}.} \bibinfo{year}{2020}\natexlab{c}.
\newblock \showarticletitle{Distance-Dependent EHMIs for the Interaction Between Automated Vehicles and Pedestrians}. In \bibinfo{booktitle}{\emph{12th International Conference on Automotive User Interfaces and Interactive Vehicular Applications}} (Virtual Event, DC, USA) \emph{(\bibinfo{series}{AutomotiveUI '20})}. \bibinfo{publisher}{ACM}, \bibinfo{address}{New York, NY, USA}, \bibinfo{pages}{192–204}.
\newblock
\showISBNx{9781450380652}
\href{https://doi.org/10.1145/3409120.3410642}{doi:\nolinkurl{10.1145/3409120.3410642}}


\bibitem[Dey et~al\mbox{.}(2018)]%
        {dey2018interface}
\bibfield{author}{\bibinfo{person}{Debargha Dey}, \bibinfo{person}{Marieke Martens}, \bibinfo{person}{Chao Wang}, \bibinfo{person}{Felix Ros}, {and} \bibinfo{person}{Jacques Terken}.} \bibinfo{year}{2018}\natexlab{}.
\newblock \showarticletitle{Interface Concepts for Intent Communication from Autonomous Vehicles to Vulnerable Road Users}. In \bibinfo{booktitle}{\emph{Adjunct Proceedings of the 10th International Conference on Automotive User Interfaces and Interactive Vehicular Applications}} (Toronto, ON, Canada) \emph{(\bibinfo{series}{AutomotiveUI ’18})}. \bibinfo{publisher}{ACM}, \bibinfo{address}{New York, NY, USA}, \bibinfo{pages}{82–86}.
\newblock
\showISBNx{9781450359474}
\href{https://doi.org/10.1145/3239092.3265946}{doi:\nolinkurl{10.1145/3239092.3265946}}


\bibitem[Dey et~al\mbox{.}(2024)]%
        {Dey2024multimodal}
\bibfield{author}{\bibinfo{person}{Debargha Dey}, \bibinfo{person}{Toros~Ufuk Senan}, \bibinfo{person}{Bart Hengeveld}, \bibinfo{person}{Mark Colley}, \bibinfo{person}{Azra Habibovic}, {and} \bibinfo{person}{Wendy Ju}.} \bibinfo{year}{2024}\natexlab{}.
\newblock \showarticletitle{{Multi-Modal eHMIs: The Relative Impact of Light and Sound in AV-Pedestrian Interaction}}. In \bibinfo{booktitle}{\emph{CHI Conference on Human Factors in Computing Systems}}. \bibinfo{publisher}{ACM}, \bibinfo{address}{New York, NY, USA}, \bibinfo{pages}{1--16}.
\newblock
\showISBNx{9798400703300}
\href{https://doi.org/10.1145/3613904.3642031}{doi:\nolinkurl{10.1145/3613904.3642031}}


\bibitem[Dey et~al\mbox{.}(2019)]%
        {dey2019gaze}
\bibfield{author}{\bibinfo{person}{Debargha Dey}, \bibinfo{person}{Francesco Walker}, \bibinfo{person}{Marieke Martens}, {and} \bibinfo{person}{Jacques Terken}.} \bibinfo{year}{2019}\natexlab{}.
\newblock \showarticletitle{Gaze Patterns in Pedestrian Interaction with Vehicles: Towards Effective Design of External Human-Machine Interfaces for Automated Vehicles}. In \bibinfo{booktitle}{\emph{Proceedings of the 11th International Conference on Automotive User Interfaces and Interactive Vehicular Applications}} (Utrecht, Netherlands) \emph{(\bibinfo{series}{AutomotiveUI '19})}. \bibinfo{publisher}{ACM}, \bibinfo{address}{New York, NY, USA}, \bibinfo{pages}{369–378}.
\newblock
\showISBNx{9781450368841}
\href{https://doi.org/10.1145/3342197.3344523}{doi:\nolinkurl{10.1145/3342197.3344523}}


\bibitem[Dietrich et~al\mbox{.}(2020)]%
        {dietrich2020automated}
\bibfield{author}{\bibinfo{person}{Andr{\'e} Dietrich}, \bibinfo{person}{Michael Tondera}, {and} \bibinfo{person}{Klaus Bengler}.} \bibinfo{year}{2020}\natexlab{}.
\newblock \showarticletitle{Automated vehicles in urban traffic: The effects of kinematics and eHMI on pedestrian crossing behaviour}.
\newblock \bibinfo{journal}{\emph{Advances in transportation studies}}  \bibinfo{volume}{2020} (\bibinfo{year}{2020}), \bibinfo{pages}{73--84}.
\newblock


\bibitem[Dudley et~al\mbox{.}(2019)]%
        {dudley2019crowdsourcing}
\bibfield{author}{\bibinfo{person}{John~J. Dudley}, \bibinfo{person}{Jason~T. Jacques}, {and} \bibinfo{person}{Per~Ola Kristensson}.} \bibinfo{year}{2019}\natexlab{}.
\newblock \showarticletitle{Crowdsourcing Interface Feature Design with Bayesian Optimization}. In \bibinfo{booktitle}{\emph{Proceedings of the 2019 CHI Conference on Human Factors in Computing Systems}} (Glasgow, Scotland Uk) \emph{(\bibinfo{series}{CHI '19})}. \bibinfo{publisher}{ACM}, \bibinfo{address}{New York, NY, USA}, \bibinfo{pages}{1–12}.
\newblock
\showISBNx{9781450359702}
\href{https://doi.org/10.1145/3290605.3300482}{doi:\nolinkurl{10.1145/3290605.3300482}}


\bibitem[Edelmann et~al\mbox{.}(2021)]%
        {edelmann2021cross}
\bibfield{author}{\bibinfo{person}{Aaron Edelmann}, \bibinfo{person}{Stefan St{\"u}mper}, {and} \bibinfo{person}{Tibor Petzoldt}.} \bibinfo{year}{2021}\natexlab{}.
\newblock \showarticletitle{Cross-cultural differences in the acceptance of decisions of automated vehicles}.
\newblock \bibinfo{journal}{\emph{Applied ergonomics}}  \bibinfo{volume}{92} (\bibinfo{year}{2021}), \bibinfo{pages}{103346}.
\newblock


\bibitem[Faas et~al\mbox{.}(2020)]%
        {faas2020longitudinal}
\bibfield{author}{\bibinfo{person}{Stefanie~M. Faas}, \bibinfo{person}{Andrea~C. Kao}, {and} \bibinfo{person}{Martin Baumann}.} \bibinfo{year}{2020}\natexlab{}.
\newblock \showarticletitle{A Longitudinal Video Study on Communicating Status and Intent for Self-Driving Vehicle – Pedestrian Interaction}. In \bibinfo{booktitle}{\emph{Proceedings of the 2020 CHI Conference on Human Factors in Computing Systems}} (Honolulu, HI, USA) \emph{(\bibinfo{series}{CHI ’20})}. \bibinfo{publisher}{ACM}, \bibinfo{address}{New York, NY, USA}, \bibinfo{pages}{1–14}.
\newblock
\showISBNx{9781450367080}
\href{https://doi.org/10.1145/3313831.3376484}{doi:\nolinkurl{10.1145/3313831.3376484}}


\bibitem[Fagnant and Kockelman(2015)]%
        {fagnant2015preparing}
\bibfield{author}{\bibinfo{person}{Daniel~J Fagnant} {and} \bibinfo{person}{Kara Kockelman}.} \bibinfo{year}{2015}\natexlab{}.
\newblock \showarticletitle{Preparing a nation for autonomous vehicles: opportunities, barriers and policy recommendations}.
\newblock \bibinfo{journal}{\emph{Transportation Research Part A: Policy and Practice}}  \bibinfo{volume}{77} (\bibinfo{year}{2015}), \bibinfo{pages}{167--181}.
\newblock


\bibitem[Florentine et~al\mbox{.}(2016)]%
        {RefWorks:doc:5cf7ad8de4b06bba938e0112}
\bibfield{author}{\bibinfo{person}{Evelyn Florentine}, \bibinfo{person}{Mark~Adam Ang}, \bibinfo{person}{Scott~Drew Pendleton}, \bibinfo{person}{Hans Andersen}, {and} \bibinfo{person}{Marcelo~H. Ang}.} \bibinfo{year}{2016}\natexlab{}.
\newblock \showarticletitle{Pedestrian Notification Methods in Autonomous Vehicles for Multi-Class Mobility-on-Demand Service}. In \bibinfo{booktitle}{\emph{Proceedings of the Fourth International Conference on Human Agent Interaction}} (Biopolis, Singapore) \emph{(\bibinfo{series}{HAI ’16})}. \bibinfo{publisher}{ACM}, \bibinfo{address}{New York, NY, USA}, \bibinfo{pages}{387–392}.
\newblock
\showISBNx{9781450345088}
\href{https://doi.org/10.1145/2974804.2974833}{doi:\nolinkurl{10.1145/2974804.2974833}}


\bibitem[Gui et~al\mbox{.}(2024)]%
        {gui2024shrinkable}
\bibfield{author}{\bibinfo{person}{Xinyue Gui}, \bibinfo{person}{Mikiya Kusunoki}, \bibinfo{person}{Bofei Huang}, \bibinfo{person}{Stela~Hanbyeol Seo}, \bibinfo{person}{Chia-Ming Chang}, \bibinfo{person}{Haoran Xie}, \bibinfo{person}{Manabu Tsukada}, {and} \bibinfo{person}{Takeo Igarashi}.} \bibinfo{year}{2024}\natexlab{}.
\newblock \showarticletitle{Shrinkable Arm-based eHMI on Autonomous Delivery Vehicle for Effective Communication with Other Road Users}. In \bibinfo{booktitle}{\emph{Proceedings of the 16th International Conference on Automotive User Interfaces and Interactive Vehicular Applications}}. \bibinfo{publisher}{ACM}, \bibinfo{address}{NY, New York, USA}, \bibinfo{pages}{305--316}.
\newblock


\bibitem[Gui et~al\mbox{.}(2023)]%
        {gui2023field}
\bibfield{author}{\bibinfo{person}{Xinyue Gui}, \bibinfo{person}{Koki Toda}, \bibinfo{person}{Stela~Hanbyeol Seo}, \bibinfo{person}{Felix~Martin Eckert}, \bibinfo{person}{Chia-Ming Chang}, \bibinfo{person}{Xiang'Anthony Chen}, {and} \bibinfo{person}{Takeo Igarashi}.} \bibinfo{year}{2023}\natexlab{}.
\newblock \showarticletitle{A field study on pedestrians’ thoughts toward a car with gazing eyes}. In \bibinfo{booktitle}{\emph{Extended Abstracts of the 2023 CHI Conference on Human Factors in Computing Systems}}. \bibinfo{publisher}{ACM}, \bibinfo{address}{NY, New York, USA}, \bibinfo{pages}{1--7}.
\newblock


\bibitem[Haimerl et~al\mbox{.}(2022)]%
        {10.1145/3546717}
\bibfield{author}{\bibinfo{person}{Mathias Haimerl}, \bibinfo{person}{Mark Colley}, {and} \bibinfo{person}{Andreas Riener}.} \bibinfo{year}{2022}\natexlab{}.
\newblock \showarticletitle{Evaluation of Common External Communication Concepts of Automated Vehicles for People With Intellectual Disabilities}.
\newblock \bibinfo{journal}{\emph{Proc. ACM Hum.-Comput. Interact.}} \bibinfo{volume}{6}, \bibinfo{number}{MHCI}, Article \bibinfo{articleno}{182} (\bibinfo{date}{Sept.} \bibinfo{year}{2022}), \bibinfo{numpages}{19}~pages.
\newblock
\href{https://doi.org/10.1145/3546717}{doi:\nolinkurl{10.1145/3546717}}


\bibitem[Hart and Staveland(1988)]%
        {hart1988development}
\bibfield{author}{\bibinfo{person}{Sandra~G Hart} {and} \bibinfo{person}{Lowell~E Staveland}.} \bibinfo{year}{1988}\natexlab{}.
\newblock \showarticletitle{Development of NASA-TLX (Task Load Index): Results of empirical and theoretical research}.
\newblock In \bibinfo{booktitle}{\emph{Advances in psychology}}. Vol.~\bibinfo{volume}{52}. \bibinfo{publisher}{Elsevier}, \bibinfo{address}{Amsterdam, The Netherlands}, \bibinfo{pages}{139--183}.
\newblock


\bibitem[Holl\"{a}nder et~al\mbox{.}(2021)]%
        {hollander2021taxonomy}
\bibfield{author}{\bibinfo{person}{Kai Holl\"{a}nder}, \bibinfo{person}{Mark Colley}, \bibinfo{person}{Enrico Rukzio}, {and} \bibinfo{person}{Andreas Butz}.} \bibinfo{year}{2021}\natexlab{}.
\newblock \showarticletitle{A Taxonomy of Vulnerable Road Users for HCI Based On A Systematic Literature Review}. In \bibinfo{booktitle}{\emph{Proceedings of the 2021 CHI Conference on Human Factors in Computing Systems}}. \bibinfo{publisher}{ACM}, \bibinfo{address}{New York, NY, USA}, Article \bibinfo{articleno}{158}, \bibinfo{numpages}{13}~pages.
\newblock
\showISBNx{9781450380966}
\urldef\tempurl%
\url{https://doi.org/10.1145/3411764.3445480}
\showURL{%
\tempurl}


\bibitem[Holl\"{a}nder et~al\mbox{.}(2020)]%
        {hollander2020smombies}
\bibfield{author}{\bibinfo{person}{Kai Holl\"{a}nder}, \bibinfo{person}{Andy Kr\"{u}ger}, {and} \bibinfo{person}{Andreas Butz}.} \bibinfo{year}{2020}\natexlab{}.
\newblock \showarticletitle{Save the Smombies: App-Assisted Street Crossing}. In \bibinfo{booktitle}{\emph{22nd International Conference on Human-Computer Interaction with Mobile Devices and Services}} (Oldenburg, Germany) \emph{(\bibinfo{series}{MobileHCI '20})}. \bibinfo{publisher}{ACM}, \bibinfo{address}{New York, NY, USA}, Article \bibinfo{articleno}{22}, \bibinfo{numpages}{11}~pages.
\newblock
\showISBNx{9781450375160}
\href{https://doi.org/10.1145/3379503.3403547}{doi:\nolinkurl{10.1145/3379503.3403547}}


\bibitem[Holl\"{a}nder et~al\mbox{.}(2019)]%
        {hollander2019overtrust}
\bibfield{author}{\bibinfo{person}{Kai Holl\"{a}nder}, \bibinfo{person}{Philipp Wintersberger}, {and} \bibinfo{person}{Andreas Butz}.} \bibinfo{year}{2019}\natexlab{}.
\newblock \showarticletitle{Overtrust in External Cues of Automated Vehicles: An Experimental Investigation}. In \bibinfo{booktitle}{\emph{Proceedings of the 11th International Conference on Automotive User Interfaces and Interactive Vehicular Applications}} (Utrecht, Netherlands) \emph{(\bibinfo{series}{AutomotiveUI ’19})}. \bibinfo{publisher}{ACM}, \bibinfo{address}{New York, NY, USA}, \bibinfo{pages}{211–221}.
\newblock
\showISBNx{9781450368841}
\href{https://doi.org/10.1145/3342197.3344528}{doi:\nolinkurl{10.1145/3342197.3344528}}


\bibitem[Hou et~al\mbox{.}(2020)]%
        {hou2020autonomous}
\bibfield{author}{\bibinfo{person}{Ming Hou}, \bibinfo{person}{Karthik Mahadevan}, \bibinfo{person}{Sowmya Somanath}, \bibinfo{person}{Ehud Sharlin}, {and} \bibinfo{person}{Lora Oehlberg}.} \bibinfo{year}{2020}\natexlab{}.
\newblock \showarticletitle{Autonomous Vehicle-Cyclist Interaction: Peril and Promise}. In \bibinfo{booktitle}{\emph{Proceedings of the 2020 CHI Conference on Human Factors in Computing Systems}} (Honolulu, HI, USA) \emph{(\bibinfo{series}{CHI ’20})}. \bibinfo{publisher}{ACM}, \bibinfo{address}{New York, NY, USA}, \bibinfo{pages}{1–12}.
\newblock
\showISBNx{9781450367080}
\href{https://doi.org/10.1145/3313831.3376884}{doi:\nolinkurl{10.1145/3313831.3376884}}


\bibitem[Kadner et~al\mbox{.}(2021)]%
        {kadner2021adaptifont}
\bibfield{author}{\bibinfo{person}{Florian Kadner}, \bibinfo{person}{Yannik Keller}, {and} \bibinfo{person}{Constantin Rothkopf}.} \bibinfo{year}{2021}\natexlab{}.
\newblock \showarticletitle{AdaptiFont: Increasing Individuals’ Reading Speed with a Generative Font Model and Bayesian Optimization}. In \bibinfo{booktitle}{\emph{Proceedings of the 2021 CHI Conference on Human Factors in Computing Systems}} (Yokohama, Japan) \emph{(\bibinfo{series}{CHI '21})}. \bibinfo{publisher}{ACM}, \bibinfo{address}{New York, NY, USA}, Article \bibinfo{articleno}{585}, \bibinfo{numpages}{11}~pages.
\newblock
\showISBNx{9781450380966}
\href{https://doi.org/10.1145/3411764.3445140}{doi:\nolinkurl{10.1145/3411764.3445140}}


\bibitem[K{\"o}rber(2019)]%
        {korber2018theoretical}
\bibfield{author}{\bibinfo{person}{Moritz K{\"o}rber}.} \bibinfo{year}{2019}\natexlab{}.
\newblock \showarticletitle{Theoretical Considerations and Development of a Questionnaire to Measure Trust in Automation}. In \bibinfo{booktitle}{\emph{Proceedings of the 20th Congress of the International Ergonomics Association (IEA 2018)}}, \bibfield{editor}{\bibinfo{person}{Sebastiano Bagnara}, \bibinfo{person}{Riccardo Tartaglia}, \bibinfo{person}{Sara Albolino}, \bibinfo{person}{Thomas Alexander}, {and} \bibinfo{person}{Yushi Fujita}} (Eds.). \bibinfo{publisher}{Springer International Publishing}, \bibinfo{address}{Cham}, \bibinfo{pages}{13--30}.
\newblock
\showISBNx{978-3-319-96074-6}


\bibitem[Koyama and Goto(2022)]%
        {koyama2022boassistant}
\bibfield{author}{\bibinfo{person}{Yuki Koyama} {and} \bibinfo{person}{Masataka Goto}.} \bibinfo{year}{2022}\natexlab{}.
\newblock \showarticletitle{BO as Assistant: Using Bayesian Optimization for Asynchronously Generating Design Suggestions}. In \bibinfo{booktitle}{\emph{Proceedings of the 35th Annual ACM Symposium on User Interface Software and Technology}} (Bend, OR, USA) \emph{(\bibinfo{series}{UIST '22})}. \bibinfo{publisher}{ACM}, \bibinfo{address}{New York, NY, USA}, Article \bibinfo{articleno}{77}, \bibinfo{numpages}{14}~pages.
\newblock
\showISBNx{9781450393201}
\href{https://doi.org/10.1145/3526113.3545664}{doi:\nolinkurl{10.1145/3526113.3545664}}


\bibitem[Koyama et~al\mbox{.}(2020)]%
        {koyama2020sequential}
\bibfield{author}{\bibinfo{person}{Yuki Koyama}, \bibinfo{person}{Issei Sato}, {and} \bibinfo{person}{Masataka Goto}.} \bibinfo{year}{2020}\natexlab{}.
\newblock \showarticletitle{Sequential Gallery for Interactive Visual Design Optimization}.
\newblock \bibinfo{journal}{\emph{ACM Trans. Graph.}} \bibinfo{volume}{39}, \bibinfo{number}{4}, Article \bibinfo{articleno}{88} (\bibinfo{date}{aug} \bibinfo{year}{2020}), \bibinfo{numpages}{12}~pages.
\newblock
\showISSN{0730-0301}
\href{https://doi.org/10.1145/3386569.3392444}{doi:\nolinkurl{10.1145/3386569.3392444}}


\bibitem[Lanzer et~al\mbox{.}(2020)]%
        {lanzer2020designing}
\bibfield{author}{\bibinfo{person}{Mirjam Lanzer}, \bibinfo{person}{Franziska Babel}, \bibinfo{person}{Fei Yan}, \bibinfo{person}{Bihan Zhang}, \bibinfo{person}{Fang You}, \bibinfo{person}{Jianmin Wang}, {and} \bibinfo{person}{Martin Baumann}.} \bibinfo{year}{2020}\natexlab{}.
\newblock \showarticletitle{Designing Communication Strategies of Autonomous Vehicles with Pedestrians: An Intercultural Study}. In \bibinfo{booktitle}{\emph{12th International Conference on Automotive User Interfaces and Interactive Vehicular Applications}} (Virtual Event, DC, USA) \emph{(\bibinfo{series}{AutomotiveUI '20})}. \bibinfo{publisher}{ACM}, \bibinfo{address}{New York, NY, USA}, \bibinfo{pages}{122–131}.
\newblock
\showISBNx{9781450380652}
\href{https://doi.org/10.1145/3409120.3410653}{doi:\nolinkurl{10.1145/3409120.3410653}}


\bibitem[Lee and Wagenmakers(2013)]%
        {lee2013bayesian}
\bibfield{author}{\bibinfo{person}{M.D. Lee} {and} \bibinfo{person}{E.J. Wagenmakers}.} \bibinfo{year}{2013}\natexlab{}.
\newblock \bibinfo{booktitle}{\emph{Bayesian Cognitive Modeling: A Practical Course}}.
\newblock \bibinfo{publisher}{Cambridge University Press}.
\newblock
\showISBNx{9781107018457}
\showLCCN{2014415414}
\urldef\tempurl%
\url{https://books.google.de/books?id=50tkAgAAQBAJ}
\showURL{%
\tempurl}


\bibitem[Lee et~al\mbox{.}(2021)]%
        {lee2021road}
\bibfield{author}{\bibinfo{person}{Yee~Mun Lee}, \bibinfo{person}{Ruth Madigan}, \bibinfo{person}{Oscar Giles}, \bibinfo{person}{Laura Garach-Morcillo}, \bibinfo{person}{Gustav Markkula}, \bibinfo{person}{Charles Fox}, \bibinfo{person}{Fanta Camara}, \bibinfo{person}{Markus Rothmueller}, \bibinfo{person}{Signe~Alexandra Vendelbo-Larsen}, \bibinfo{person}{Pernille~Holm Rasmussen}, {et~al\mbox{.}}} \bibinfo{year}{2021}\natexlab{}.
\newblock \showarticletitle{Road users rarely use explicit communication when interacting in today’s traffic: implications for automated vehicles}.
\newblock \bibinfo{journal}{\emph{Cognition, Technology \& Work}}  \bibinfo{volume}{23} (\bibinfo{year}{2021}), \bibinfo{pages}{367--380}.
\newblock


\bibitem[Liao(2023)]%
        {liao2023human}
\bibfield{author}{\bibinfo{person}{Yi-Chi Liao}.} \bibinfo{year}{2023}\natexlab{}.
\newblock \emph{\bibinfo{title}{Human-in-the-Loop Design Optimization}}.
\newblock \bibinfo{thesistype}{Ph.\,D. Dissertation}. \bibinfo{school}{Aalto University}.
\newblock
\urldef\tempurl%
\url{https://urn.fi/URN:ISBN:978-952-64-1580-2}
\showURL{%
\tempurl}


\bibitem[Liao et~al\mbox{.}(2023)]%
        {liao2023interaction}
\bibfield{author}{\bibinfo{person}{Yi-Chi Liao}, \bibinfo{person}{John~J Dudley}, \bibinfo{person}{George~B Mo}, \bibinfo{person}{Chun-Lien Cheng}, \bibinfo{person}{Liwei Chan}, \bibinfo{person}{Antti Oulasvirta}, {and} \bibinfo{person}{Per~Ola Kristensson}.} \bibinfo{year}{2023}\natexlab{}.
\newblock \showarticletitle{Interaction Design With Multi-objective Bayesian Optimization}.
\newblock \bibinfo{journal}{\emph{IEEE Pervasive Computing}} \bibinfo{volume}{22}, \bibinfo{number}{1} (\bibinfo{year}{2023}), \bibinfo{pages}{29--38}.
\newblock


\bibitem[L{\"o}cken et~al\mbox{.}(2020)]%
        {locken2020wecare}
\bibfield{author}{\bibinfo{person}{Andreas L{\"o}cken}, \bibinfo{person}{Mark Colley}, \bibinfo{person}{Andrii Matviienko}, \bibinfo{person}{Kai Holl{\"a}nder}, \bibinfo{person}{Debargha Dey}, \bibinfo{person}{Azra Habibovic}, \bibinfo{person}{Andrew~L Kun}, \bibinfo{person}{Susanne Boll}, {and} \bibinfo{person}{Andreas Riener}.} \bibinfo{year}{2020}\natexlab{}.
\newblock \showarticletitle{WeCARe: Workshop on Inclusive Communication between Automated Vehicles and Vulnerable Road Users}. In \bibinfo{booktitle}{\emph{22nd International Conference on Human-Computer Interaction with Mobile Devices and Services}}. \bibinfo{publisher}{ACM}, \bibinfo{address}{New York, NY, USA}, \bibinfo{pages}{1--5}.
\newblock


\bibitem[L\"{o}cken et~al\mbox{.}(2019)]%
        {locken2019should}
\bibfield{author}{\bibinfo{person}{Andreas L\"{o}cken}, \bibinfo{person}{Carmen Golling}, {and} \bibinfo{person}{Andreas Riener}.} \bibinfo{year}{2019}\natexlab{}.
\newblock \showarticletitle{How Should Automated Vehicles Interact with Pedestrians? A Comparative Analysis of Interaction Concepts in Virtual Reality}. In \bibinfo{booktitle}{\emph{Proceedings of the 11th International Conference on Automotive User Interfaces and Interactive Vehicular Applications}} (Utrecht, Netherlands) \emph{(\bibinfo{series}{AutomotiveUI ’19})}. \bibinfo{publisher}{ACM}, \bibinfo{address}{New York, NY, USA}, \bibinfo{pages}{262–274}.
\newblock
\showISBNx{9781450368841}
\href{https://doi.org/10.1145/3342197.3344544}{doi:\nolinkurl{10.1145/3342197.3344544}}


\bibitem[Lundgren et~al\mbox{.}(2017)]%
        {RefWorks:doc:5cf8aa47e4b006bc06a90b0a}
\bibfield{author}{\bibinfo{person}{Victor~Malmsten Lundgren}, \bibinfo{person}{Azra Habibovic}, \bibinfo{person}{Jonas Andersson}, \bibinfo{person}{Tobias Lagstr{\"o}m}, \bibinfo{person}{Maria Nilsson}, \bibinfo{person}{Anna Sirkka}, \bibinfo{person}{Johan Fagerl{\"o}nn}, \bibinfo{person}{Rikard Fredriksson}, \bibinfo{person}{Claes Edgren}, \bibinfo{person}{Stas Krupenia}, {and} \bibinfo{person}{Dennis Salu{\"a}{\"a}r}.} \bibinfo{year}{2017}\natexlab{}.
\newblock \showarticletitle{Will There Be New Communication Needs When Introducing Automated Vehicles to the Urban Context?}. In \bibinfo{booktitle}{\emph{Advances in Human Aspects of Transportation}}, \bibfield{editor}{\bibinfo{person}{Neville~A. Stanton}, \bibinfo{person}{Steven Landry}, \bibinfo{person}{Giuseppe Di~Bucchianico}, {and} \bibinfo{person}{Andrea Vallicelli}} (Eds.). \bibinfo{publisher}{Springer International Publishing}, \bibinfo{address}{Cham}, \bibinfo{pages}{485--497}.
\newblock
\showISBNx{978-3-319-41682-3}


\bibitem[Mahadevan et~al\mbox{.}(2018)]%
        {mahadevan2018communicating}
\bibfield{author}{\bibinfo{person}{Karthik Mahadevan}, \bibinfo{person}{Sowmya Somanath}, {and} \bibinfo{person}{Ehud Sharlin}.} \bibinfo{year}{2018}\natexlab{}.
\newblock \showarticletitle{Communicating Awareness and Intent in Autonomous Vehicle-Pedestrian Interaction}. In \bibinfo{booktitle}{\emph{Proceedings of the 2018 CHI Conference on Human Factors in Computing Systems}} (Montreal QC, Canada) \emph{(\bibinfo{series}{CHI ’18})}. \bibinfo{publisher}{ACM}, \bibinfo{address}{New York, NY, USA}, \bibinfo{pages}{1–12}.
\newblock
\showISBNx{9781450356206}
\href{https://doi.org/10.1145/3173574.3174003}{doi:\nolinkurl{10.1145/3173574.3174003}}


\bibitem[Marler and Arora(2004)]%
        {marler_survey_2004}
\bibfield{author}{\bibinfo{person}{R.T. Marler} {and} \bibinfo{person}{J.S. Arora}.} \bibinfo{year}{2004}\natexlab{}.
\newblock \showarticletitle{Survey of multi-objective optimization methods for engineering}.
\newblock \bibinfo{journal}{\emph{Structural and Multidisciplinary Optimization}} \bibinfo{volume}{26}, \bibinfo{number}{6} (\bibinfo{date}{April} \bibinfo{year}{2004}), \bibinfo{pages}{369--395}.
\newblock
\showISSN{1615-147X, 1615-1488}
\href{https://doi.org/10.1007/s00158-003-0368-6}{doi:\nolinkurl{10.1007/s00158-003-0368-6}}


\bibitem[Mersmann(2024)]%
        {emoa}
\bibfield{author}{\bibinfo{person}{Olaf Mersmann}.} \bibinfo{year}{2024}\natexlab{}.
\newblock \bibinfo{booktitle}{\emph{emoa: Evolutionary Multiobjective Optimization Algorithms}}.
\newblock
\href{https://doi.org/10.32614/CRAN.package.emoa}{doi:\nolinkurl{10.32614/CRAN.package.emoa}}
\newblock
\shownote{R package version 0.5-3}.


\bibitem[Moore et~al\mbox{.}(2019)]%
        {moore2019case}
\bibfield{author}{\bibinfo{person}{Dylan Moore}, \bibinfo{person}{Rebecca Currano}, \bibinfo{person}{G.~Ella Strack}, {and} \bibinfo{person}{David Sirkin}.} \bibinfo{year}{2019}\natexlab{}.
\newblock \showarticletitle{The Case for Implicit External Human-Machine Interfaces for Autonomous Vehicles}. In \bibinfo{booktitle}{\emph{Proceedings of the 11th International Conference on Automotive User Interfaces and Interactive Vehicular Applications}} (Utrecht, Netherlands) \emph{(\bibinfo{series}{AutomotiveUI ’19})}. \bibinfo{publisher}{ACM}, \bibinfo{address}{New York, NY, USA}, \bibinfo{pages}{295–307}.
\newblock
\showISBNx{9781450368841}
\href{https://doi.org/10.1145/3342197.3345320}{doi:\nolinkurl{10.1145/3342197.3345320}}


\bibitem[Nguyen et~al\mbox{.}(2019)]%
        {nguyen2019designing}
\bibfield{author}{\bibinfo{person}{Trung~Thanh Nguyen}, \bibinfo{person}{Kai Holl\"{a}nder}, \bibinfo{person}{Marius Hoggenmueller}, \bibinfo{person}{Callum Parker}, {and} \bibinfo{person}{Martin Tomitsch}.} \bibinfo{year}{2019}\natexlab{}.
\newblock \showarticletitle{Designing for Projection-Based Communication between Autonomous Vehicles and Pedestrians}. In \bibinfo{booktitle}{\emph{Proceedings of the 11th International Conference on Automotive User Interfaces and Interactive Vehicular Applications}} (Utrecht, Netherlands) \emph{(\bibinfo{series}{AutomotiveUI ’19})}. \bibinfo{publisher}{ACM}, \bibinfo{address}{New York, NY, USA}, \bibinfo{pages}{284–294}.
\newblock
\showISBNx{9781450368841}
\href{https://doi.org/10.1145/3342197.3344543}{doi:\nolinkurl{10.1145/3342197.3344543}}


\bibitem[Normark(2015)]%
        {normark2015design}
\bibfield{author}{\bibinfo{person}{Carl~Jörgen Normark}.} \bibinfo{year}{2015}\natexlab{}.
\newblock \showarticletitle{Design and Evaluation of a Touch-Based Personalizable In-Vehicle User Interface}.
\newblock \bibinfo{journal}{\emph{International Journal of Human–Computer Interaction}} \bibinfo{volume}{31}, \bibinfo{number}{11} (\bibinfo{year}{2015}), \bibinfo{pages}{731--745}.
\newblock
\href{https://doi.org/10.1080/10447318.2015.1045240}{doi:\nolinkurl{10.1080/10447318.2015.1045240}}
\showeprint{https://doi.org/10.1080/10447318.2015.1045240}


\bibitem[O'Dowd and Pollet(2018)]%
        {o2018gender}
\bibfield{author}{\bibinfo{person}{Eryn O'Dowd} {and} \bibinfo{person}{Thomas~V Pollet}.} \bibinfo{year}{2018}\natexlab{}.
\newblock \showarticletitle{Gender differences in use of a pedestrian crossing: an observational study in newcastle upon tyne}.
\newblock \bibinfo{journal}{\emph{Letters on Evolutionary Behavioral Science}} \bibinfo{volume}{9}, \bibinfo{number}{1} (\bibinfo{year}{2018}), \bibinfo{pages}{1--4}.
\newblock


\bibitem[Oudshoorn et~al\mbox{.}(2021)]%
        {oudshoorn2021bio}
\bibfield{author}{\bibinfo{person}{Max Oudshoorn}, \bibinfo{person}{Joost de Winter}, \bibinfo{person}{Pavlo Bazilinskyy}, {and} \bibinfo{person}{Dimitra Dodou}.} \bibinfo{year}{2021}\natexlab{}.
\newblock \showarticletitle{Bio-inspired intent communication for automated vehicles}.
\newblock \bibinfo{journal}{\emph{Transportation research part F: traffic psychology and behaviour}}  \bibinfo{volume}{80} (\bibinfo{year}{2021}), \bibinfo{pages}{127--140}.
\newblock


\bibitem[Rasouli et~al\mbox{.}(2017)]%
        {rasouli2017understanding}
\bibfield{author}{\bibinfo{person}{Amir Rasouli}, \bibinfo{person}{Iuliia Kotseruba}, {and} \bibinfo{person}{John~K Tsotsos}.} \bibinfo{year}{2017}\natexlab{}.
\newblock \showarticletitle{Understanding pedestrian behavior in complex traffic scenes}.
\newblock \bibinfo{journal}{\emph{IEEE Transactions on Intelligent Vehicles}} \bibinfo{volume}{3}, \bibinfo{number}{1} (\bibinfo{year}{2017}), \bibinfo{pages}{61--70}.
\newblock


\bibitem[Sadeghian et~al\mbox{.}(2020)]%
        {sadeghian2020exploration}
\bibfield{author}{\bibinfo{person}{Shadan Sadeghian}, \bibinfo{person}{Marc Hassenzahl}, {and} \bibinfo{person}{Kai Eckoldt}.} \bibinfo{year}{2020}\natexlab{}.
\newblock \showarticletitle{An Exploration of Prosocial Aspects of Communication Cues between Automated Vehicles and Pedestrians}. In \bibinfo{booktitle}{\emph{12th International Conference on Automotive User Interfaces and Interactive Vehicular Applications}} (Virtual Event, DC, USA) \emph{(\bibinfo{series}{AutomotiveUI '20})}. \bibinfo{publisher}{ACM}, \bibinfo{address}{New York, NY, USA}, \bibinfo{pages}{205–211}.
\newblock
\showISBNx{9781450380652}
\href{https://doi.org/10.1145/3409120.3410657}{doi:\nolinkurl{10.1145/3409120.3410657}}


\bibitem[Sahin et~al\mbox{.}(2021)]%
        {sahin2021workshop}
\bibfield{author}{\bibinfo{person}{Hatice Sahin}, \bibinfo{person}{Heiko Mueller}, \bibinfo{person}{Shadan Sadeghian}, \bibinfo{person}{Debargha Dey}, \bibinfo{person}{Andreas L\"{o}cken}, \bibinfo{person}{Andrii Matviienko}, \bibinfo{person}{Mark Colley}, \bibinfo{person}{Azra Habibovic}, {and} \bibinfo{person}{Philipp Wintersberger}.} \bibinfo{year}{2021}\natexlab{}.
\newblock \showarticletitle{Workshop on Prosocial Behavior in Future Mixed Traffic}. In \bibinfo{booktitle}{\emph{13th International Conference on Automotive User Interfaces and Interactive Vehicular Applications}}. \bibinfo{publisher}{ACM}, \bibinfo{address}{New York, NY, USA}, \bibinfo{pages}{167–170}.
\newblock
\showISBNx{9781450386418}
\urldef\tempurl%
\url{https://doi.org/10.1145/3473682.3477438}
\showURL{%
\tempurl}


\bibitem[Schlackl et~al\mbox{.}(2020)]%
        {schlackl2020ehmi}
\bibfield{author}{\bibinfo{person}{Dominik Schlackl}, \bibinfo{person}{Klemens Weigl}, {and} \bibinfo{person}{Andreas Riener}.} \bibinfo{year}{2020}\natexlab{}.
\newblock \showarticletitle{eHMI visualization on the entire car body: results of a comparative evaluation of concepts for the communication between AVs and manual drivers}.
\newblock In \bibinfo{booktitle}{\emph{Proceedings of Mensch und Computer 2020}}. \bibinfo{publisher}{ACM}, \bibinfo{address}{New York, NY, USA}, \bibinfo{pages}{79--83}.
\newblock


\bibitem[Sie\ss et~al\mbox{.}(2015)]%
        {RefWorks:doc:5cd92775e4b0487541989799}
\bibfield{author}{\bibinfo{person}{Andreas Sie\ss}, \bibinfo{person}{Kathleen H\"ubel}, \bibinfo{person}{Daniel Hepperle}, \bibinfo{person}{Andreas Dronov}, \bibinfo{person}{Christian Hufnagel}, \bibinfo{person}{Julia Aktun}, {and} \bibinfo{person}{Matthias W\"olfel}.} \bibinfo{year}{2015}\natexlab{}.
\newblock \showarticletitle{Hybrid City Lighting-Improving Pedestrians' Safety through Proactive Street Lighting}. In \bibinfo{booktitle}{\emph{2015 International Conference on Cyberworlds (CW)}}. \bibinfo{publisher}{IEEE}, \bibinfo{address}{New York, NY, USA}, \bibinfo{pages}{46--49}.
\newblock


\bibitem[Sobol(1967)]%
        {sobol1967}
\bibfield{author}{\bibinfo{person}{I.~M. Sobol}.} \bibinfo{year}{1967}\natexlab{}.
\newblock \showarticletitle{On the distribution of points in a cube and the approximate evaluation of integrals}.
\newblock \bibinfo{journal}{\emph{U. S. S. R. Comput. Math. and Math. Phys.}}  \bibinfo{volume}{7} (\bibinfo{year}{1967}), \bibinfo{pages}{86--112}.
\newblock


\bibitem[Sridhar et~al\mbox{.}(2015)]%
        {dridhar2015midair}
\bibfield{author}{\bibinfo{person}{Srinath Sridhar}, \bibinfo{person}{Anna~Maria Feit}, \bibinfo{person}{Christian Theobalt}, {and} \bibinfo{person}{Antti Oulasvirta}.} \bibinfo{year}{2015}\natexlab{}.
\newblock \showarticletitle{Investigating the Dexterity of Multi-Finger Input for Mid-Air Text Entry}. In \bibinfo{booktitle}{\emph{Proceedings of the 33rd Annual ACM Conference on Human Factors in Computing Systems}} (Seoul, Republic of Korea) \emph{(\bibinfo{series}{CHI '15})}. \bibinfo{publisher}{ACM}, \bibinfo{address}{New York, NY, USA}, \bibinfo{pages}{3643–3652}.
\newblock
\showISBNx{9781450331456}
\href{https://doi.org/10.1145/2702123.2702136}{doi:\nolinkurl{10.1145/2702123.2702136}}


\bibitem[Tran et~al\mbox{.}(2024a)]%
        {tran2024exploring}
\bibfield{author}{\bibinfo{person}{Tram Thi~Minh Tran}, \bibinfo{person}{Callum Parker}, \bibinfo{person}{Marius Hoggenm{\"u}ller}, \bibinfo{person}{Yiyuan Wang}, {and} \bibinfo{person}{Martin Tomitsch}.} \bibinfo{year}{2024}\natexlab{a}.
\newblock \showarticletitle{Exploring the Impact of Interconnected External Interfaces in Autonomous Vehicles on Pedestrian Safety and Experience}. In \bibinfo{booktitle}{\emph{Proceedings of the CHI Conference on Human Factors in Computing Systems}}. \bibinfo{publisher}{ACM}, \bibinfo{address}{New York, NY, USA}, \bibinfo{pages}{1--17}.
\newblock


\bibitem[Tran et~al\mbox{.}(2024b)]%
        {10.1145/3678506}
\bibfield{author}{\bibinfo{person}{Tram Thi~Minh Tran}, \bibinfo{person}{Callum Parker}, \bibinfo{person}{Xinyan Yu}, \bibinfo{person}{Debargha Dey}, \bibinfo{person}{Marieke Martens}, \bibinfo{person}{Pavlo Bazilinskyy}, {and} \bibinfo{person}{Martin Tomitsch}.} \bibinfo{year}{2024}\natexlab{b}.
\newblock \showarticletitle{Evaluating Autonomous Vehicle External Communication Using a Multi-Pedestrian VR Simulator}.
\newblock \bibinfo{journal}{\emph{Proc. ACM Interact. Mob. Wearable Ubiquitous Technol.}} \bibinfo{volume}{8}, \bibinfo{number}{3}, Article \bibinfo{articleno}{130} (\bibinfo{date}{sep} \bibinfo{year}{2024}), \bibinfo{numpages}{26}~pages.
\newblock
\href{https://doi.org/10.1145/3678506}{doi:\nolinkurl{10.1145/3678506}}


\bibitem[Van Der~Laan et~al\mbox{.}(1997)]%
        {van1997simple}
\bibfield{author}{\bibinfo{person}{Jinke~D Van Der~Laan}, \bibinfo{person}{Adriaan Heino}, {and} \bibinfo{person}{Dick De~Waard}.} \bibinfo{year}{1997}\natexlab{}.
\newblock \showarticletitle{A simple procedure for the assessment of acceptance of advanced transport telematics}.
\newblock \bibinfo{journal}{\emph{Transportation Research Part C: Emerging Technologies}} \bibinfo{volume}{5}, \bibinfo{number}{1} (\bibinfo{year}{1997}), \bibinfo{pages}{1--10}.
\newblock


\bibitem[Wobbrock and Kientz(2016)]%
        {Wobbrock.2016}
\bibfield{author}{\bibinfo{person}{Jacob~O. Wobbrock} {and} \bibinfo{person}{Julie~A. Kientz}.} \bibinfo{year}{2016}\natexlab{}.
\newblock \showarticletitle{Research Contributions in Human-Computer Interaction}.
\newblock \bibinfo{journal}{\emph{Interactions}} \bibinfo{volume}{23}, \bibinfo{number}{3} (\bibinfo{date}{apr} \bibinfo{year}{2016}), \bibinfo{pages}{38–44}.
\newblock
\showISSN{1072-5520}
\href{https://doi.org/10.1145/2907069}{doi:\nolinkurl{10.1145/2907069}}


\bibitem[Zhong et~al\mbox{.}(2021)]%
        {zhong2021spacewalker}
\bibfield{author}{\bibinfo{person}{Mingyuan Zhong}, \bibinfo{person}{Gang Li}, {and} \bibinfo{person}{Yang Li}.} \bibinfo{year}{2021}\natexlab{}.
\newblock \showarticletitle{Spacewalker: Rapid UI Design Exploration Using Lightweight Markup Enhancement and Crowd Genetic Programming}. In \bibinfo{booktitle}{\emph{Proceedings of the 2021 CHI Conference on Human Factors in Computing Systems}} (Yokohama, Japan) \emph{(\bibinfo{series}{CHI '21})}. \bibinfo{publisher}{ACM}, \bibinfo{address}{New York, NY, USA}, Article \bibinfo{articleno}{315}, \bibinfo{numpages}{11}~pages.
\newblock
\showISBNx{9781450380966}
\href{https://doi.org/10.1145/3411764.3445326}{doi:\nolinkurl{10.1145/3411764.3445326}}


\bibitem[Zimmermann and Wettach(2017)]%
        {zimmermann2017first}
\bibfield{author}{\bibinfo{person}{Raphael Zimmermann} {and} \bibinfo{person}{Reto Wettach}.} \bibinfo{year}{2017}\natexlab{}.
\newblock \showarticletitle{First Step into Visceral Interaction with Autonomous Vehicles}. In \bibinfo{booktitle}{\emph{Proceedings of the 9th International Conference on Automotive User Interfaces and Interactive Vehicular Applications}} (Oldenburg, Germany) \emph{(\bibinfo{series}{AutomotiveUI ’17})}. \bibinfo{publisher}{ACM}, \bibinfo{address}{New York, NY, USA}, \bibinfo{pages}{58–64}.
\newblock
\showISBNx{9781450351508}
\href{https://doi.org/10.1145/3122986.3122988}{doi:\nolinkurl{10.1145/3122986.3122988}}


\end{thebibliography}

\appendix

\section{Procedure --- Introduction}\label{app:introduction}

\begin{quote}
\textit{You will be crossing a road in a virtual reality (VR) environment. You will encounter automated and manually driven vehicles. Automated vehicles are identified by a strip on the upper part of the windshield. If you feel uncomfortable at any time, you can stop the study without giving any reasons and without any disadvantages. In this case, simply remove the VR goggles and inform the study leader.\\
You will be shown different scenarios, one after the other. The aim of all scenarios is to reach the checkpoint (blue) on the opposite side of the road. If an automated vehicle recognizes that you want to cross the road, it will communicate with you, depending on the passage. This communication will change over the course of the study.}
\end{quote}

\section{Results}\label{app:results}

\subsection{Objective Values over Iterations}\label{app:iterations}

\begin{figure*}[ht!]
\centering
             \begin{subfigure}[b]{0.49\linewidth}
             \includegraphics[width=\linewidth]{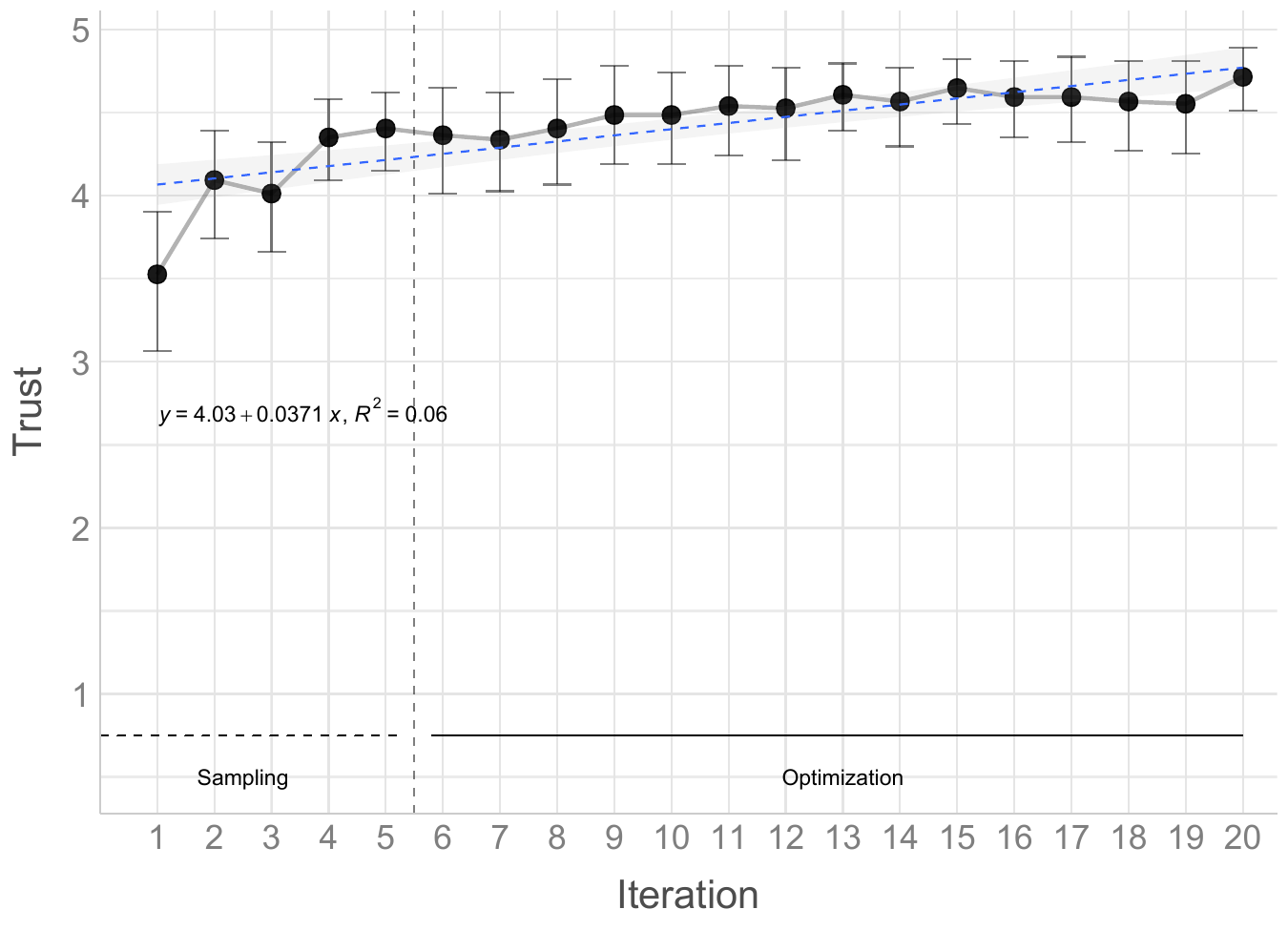}
   \caption{Progression of \textbf{trust} values over MOBO iterations.}
   \label{fig:runs_trust}
    \Description{}
              \end{subfigure}
         \begin{subfigure}[b]{0.49\linewidth}
    \includegraphics[width=\linewidth]{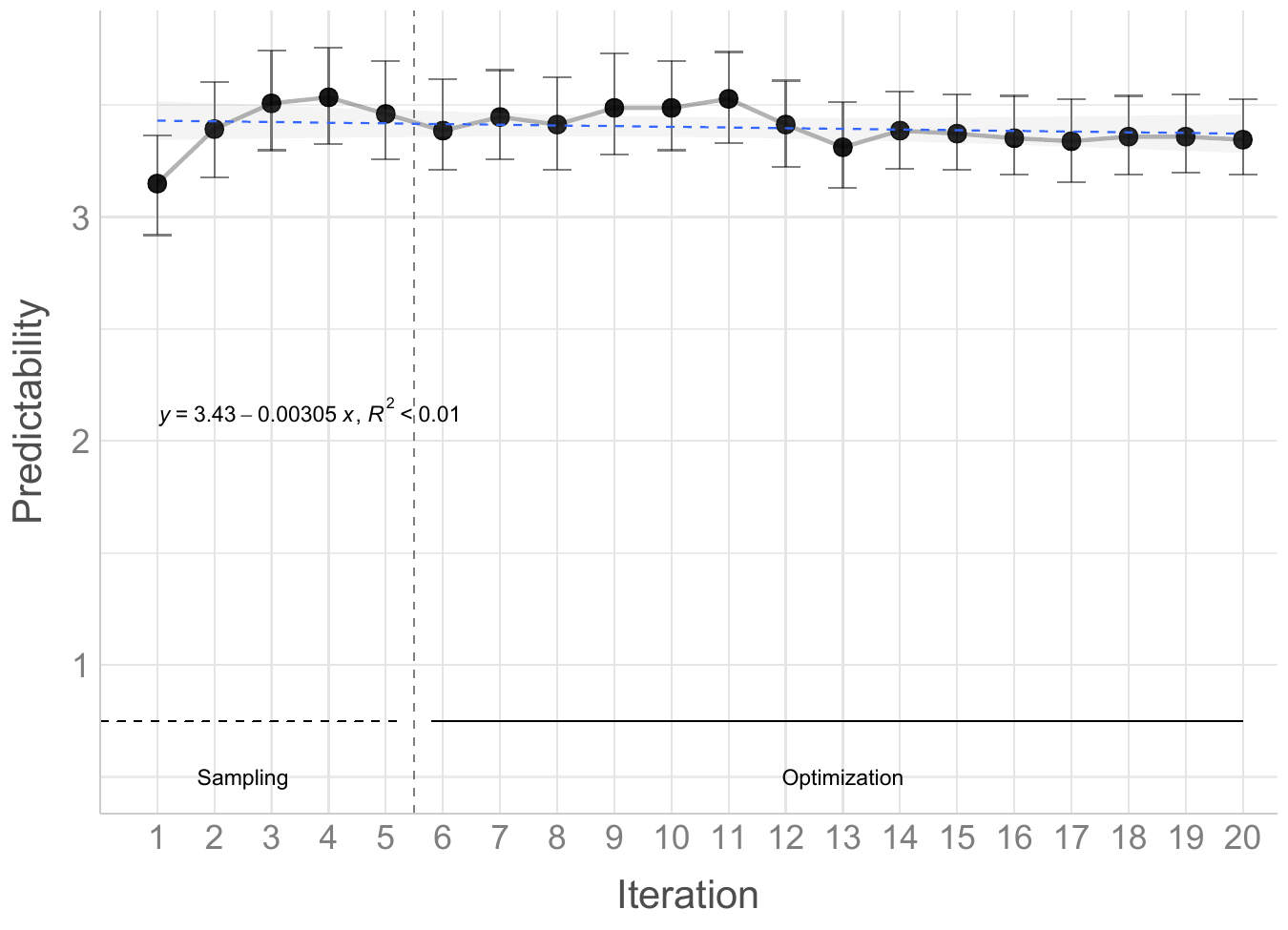}
   \caption{Progression of \textbf{predictability} values over MOBO iterations.}
   \label{fig:runs_pred}
    \Description{}
  \end{subfigure}
    \caption{Value progression of trust and predictability.}
    \Description{Progression of trust (going steadily upward) and predictability (going steadily upward) over MOBO iterations.}
\end{figure*}

\begin{figure*}[ht!]
\centering
         \begin{subfigure}[b]{0.49\linewidth}
    \includegraphics[width=\linewidth]{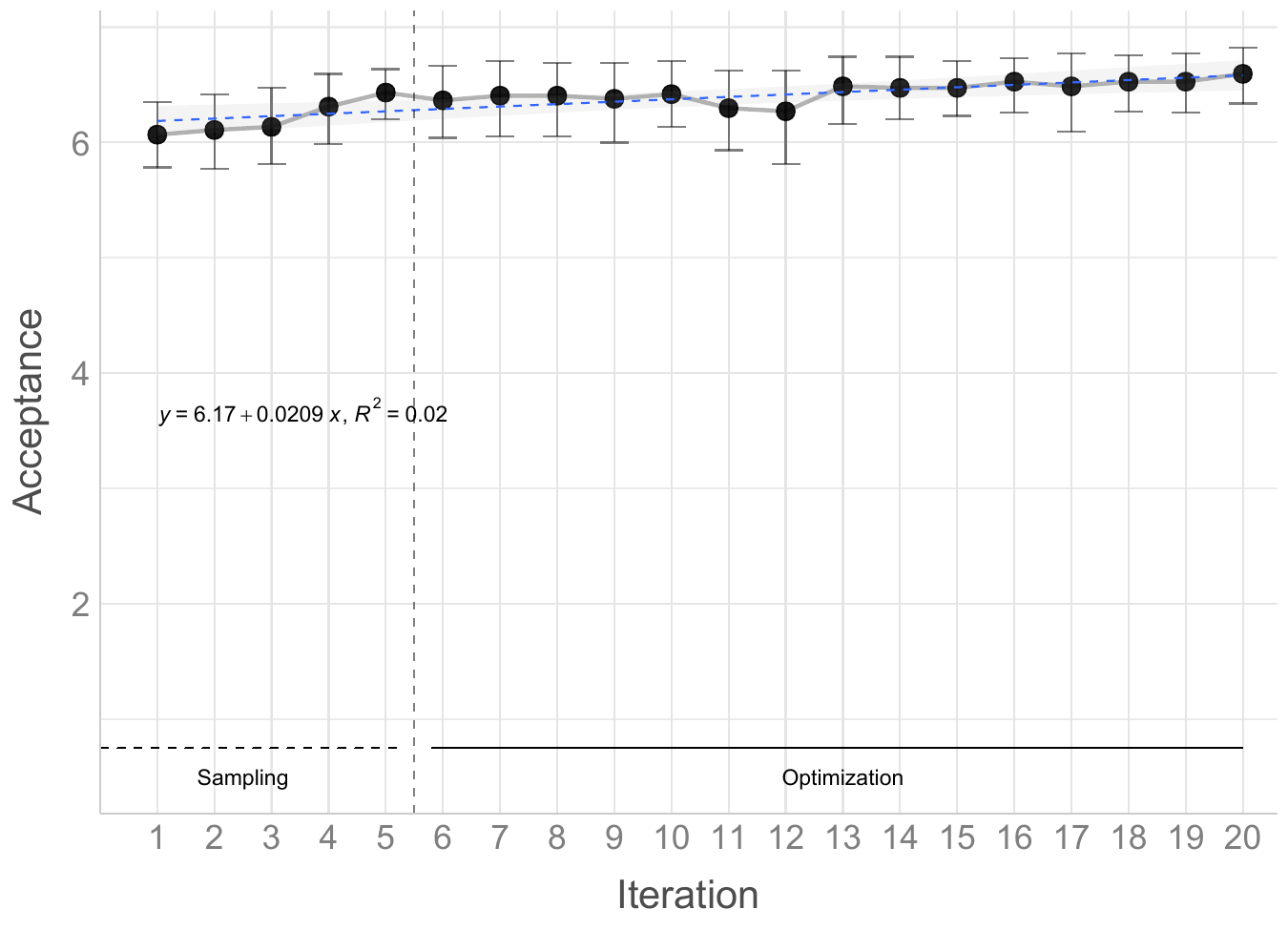}
   \caption{Progression of \textbf{acceptance} over MOBO iterations.}
   \label{fig:runs_acc}
    \Description{}
     \end{subfigure}
    \begin{subfigure}[b]{0.49\linewidth}
                          \includegraphics[width=\linewidth]{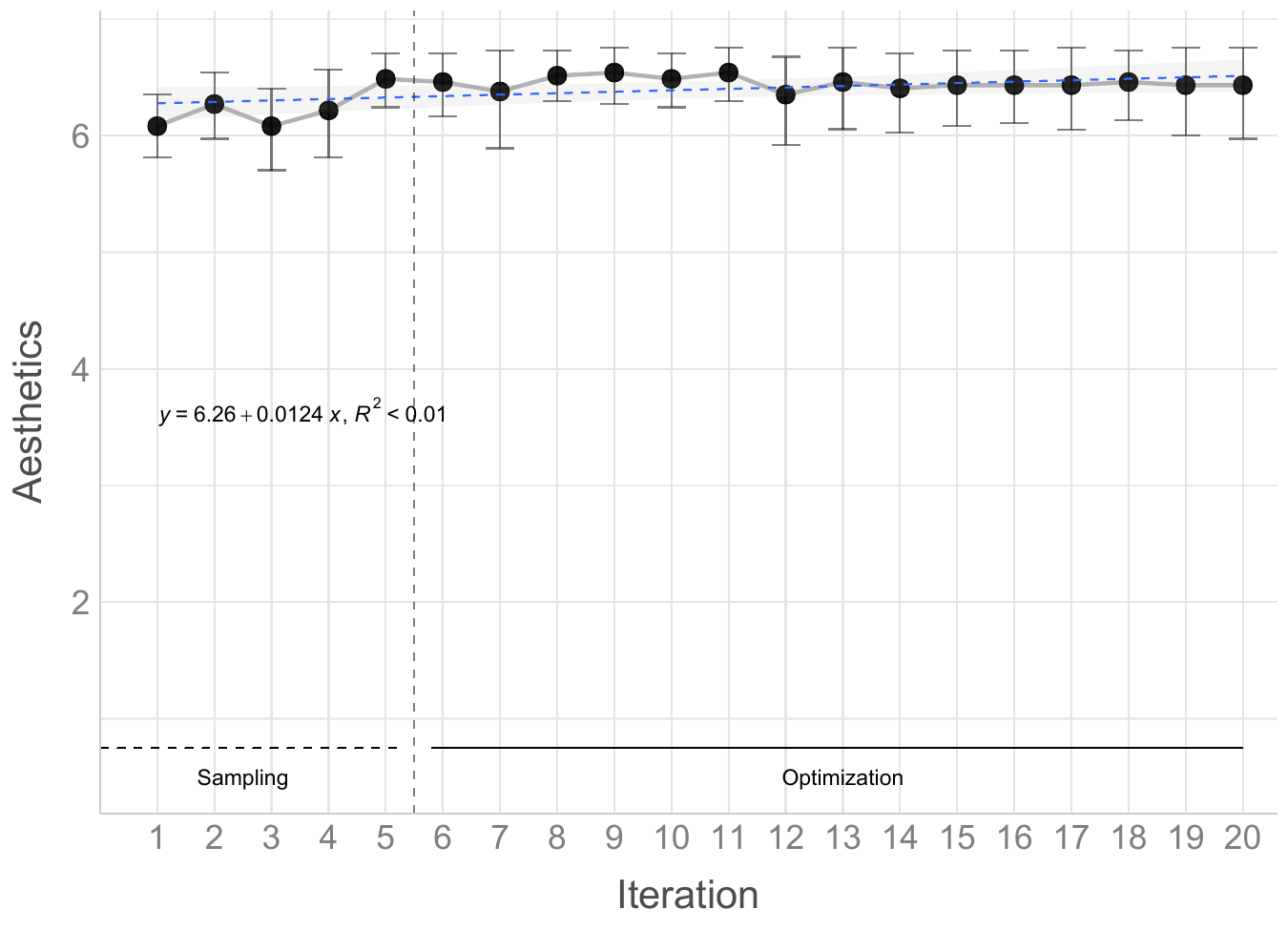}
   \caption{Progression of \textbf{aesthetics} over MOBO iterations.}
   \label{fig:runs_aes}
    \Description{}
        \end{subfigure}
    \caption{Value progression of acceptance and aesthetics. }
    \Description{Progression of acceptance (going steadily upward) and aesthetics (going steadily upward) over MOBO iterations.}
\end{figure*}

\begin{figure*}[ht!]
\centering
         \begin{subfigure}[b]{0.49\linewidth}
    \includegraphics[width=\linewidth]{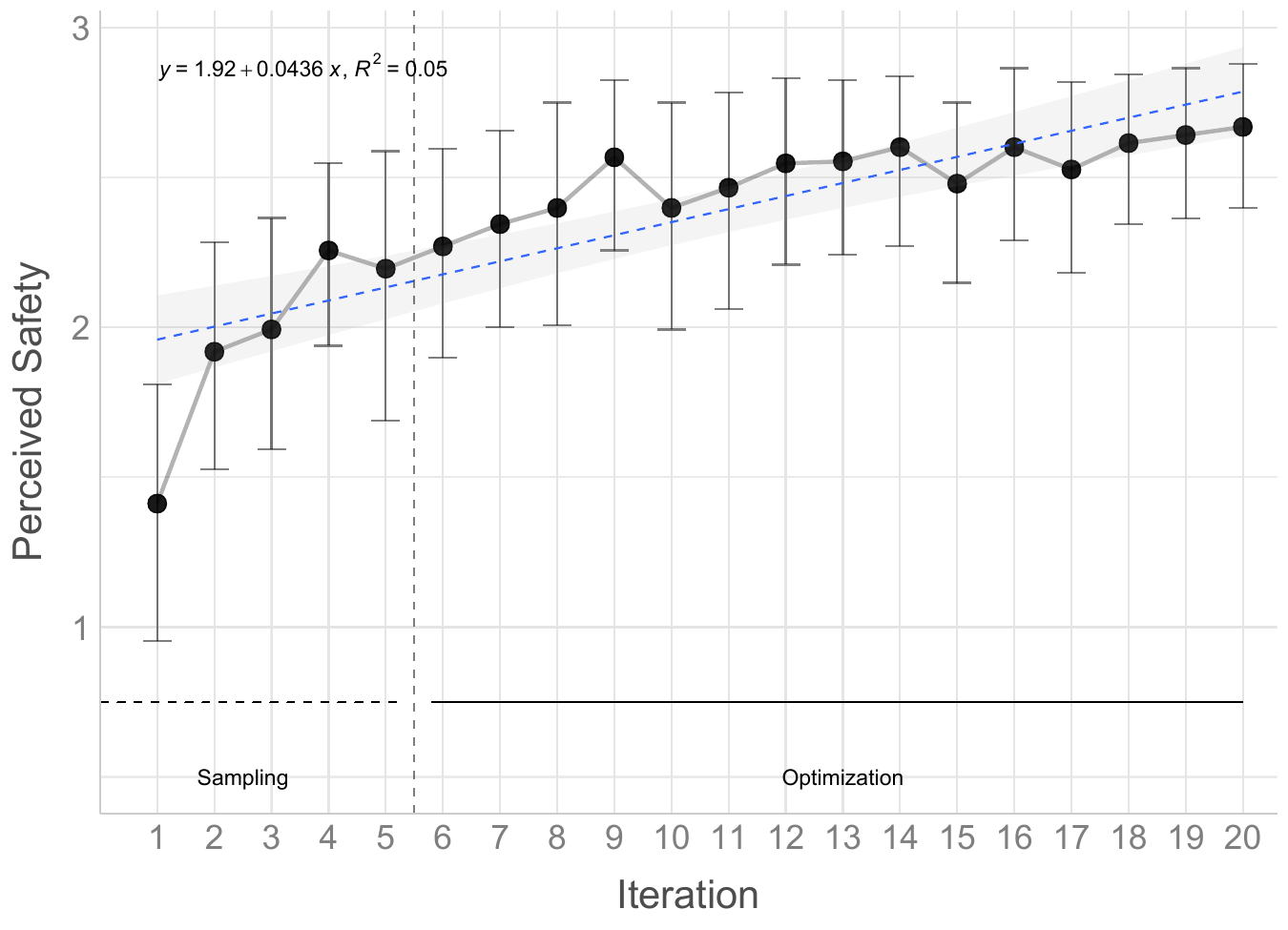}
   \caption{Progression of \textbf{perceived safety} over MOBO iterations.}
   \label{fig:runs_ps}
    \Description{}
     \end{subfigure}
         \begin{subfigure}[b]{0.49\linewidth}
                          \includegraphics[width=\linewidth]{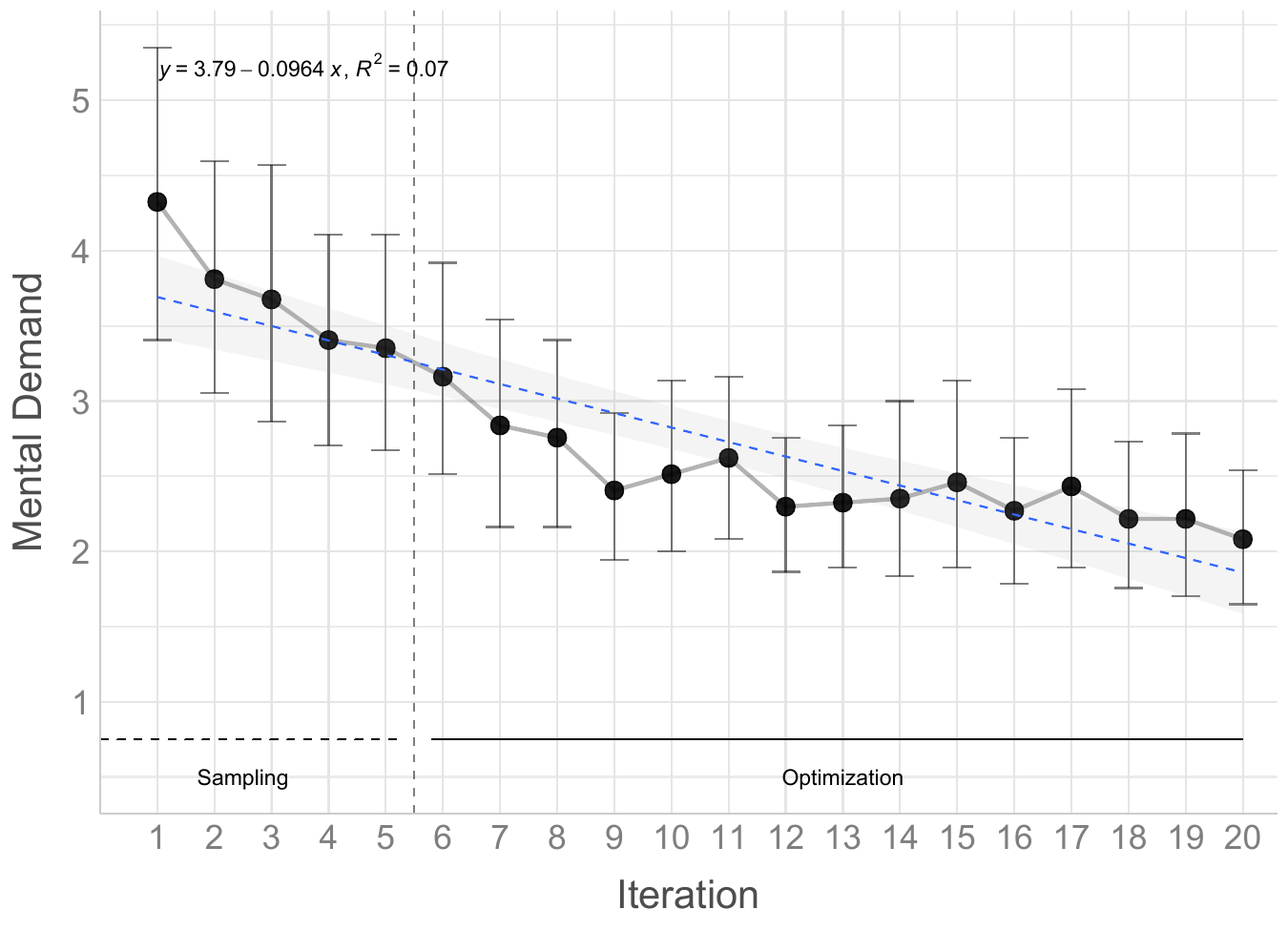}
   \caption{Progression of \textbf{mental demand} over MOBO iterations.}
   \label{fig:runs_cog}
    \Description{}
        \end{subfigure}
    \caption{Value progression of perceived safety and mental demand. }
    \Description{Progression of perceived safety (going steadily upward) and mental demand (going steadily downward) over MOBO iterations.}
\end{figure*}

\begin{figure*}[ht!]
\centering
    \includegraphics[width=0.5\linewidth]{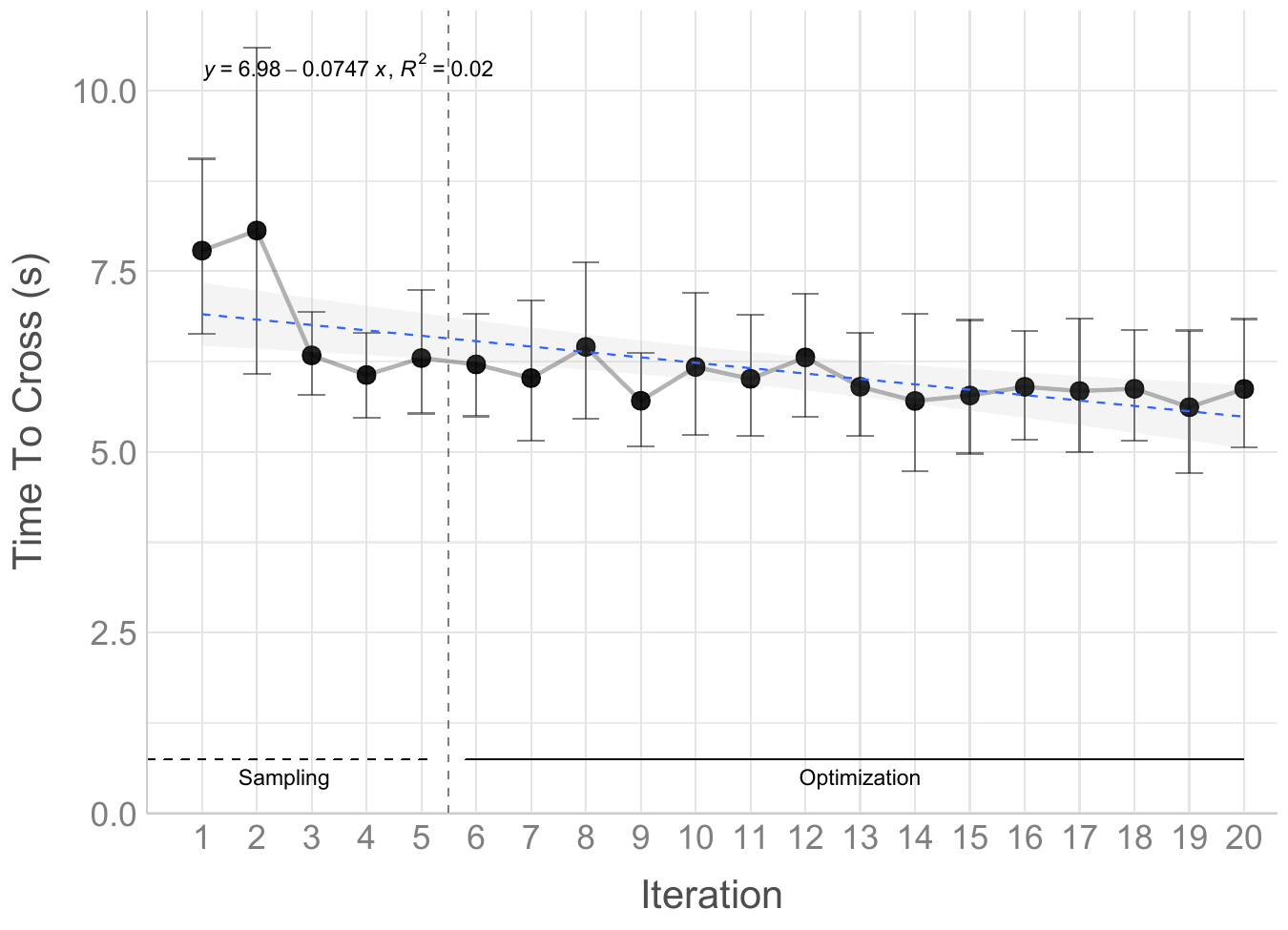}
   \caption{Progression of \textbf{Time to start crossing} over MOBO iterations.}
   \label{fig:runs_time}
    \Description{Progression of Time to start crossing (going steadily downward) over MOBO iterations.}
\end{figure*}

\end{document}